\documentclass[aps,twocolumn,prd,showpacs,showkeys,preprintnumbers,superscriptaddress,bibnotes,floatfix,longbibliography,notitlepage,nofootinbib]{revtex4-2}
\usepackage[colorlinks=true,breaklinks=true]{hyperref}
\usepackage[utf8]{inputenc}
\usepackage{url}
\usepackage{mathtools}
\usepackage[dvipsnames]{xcolor}
\definecolor{darkred}{rgb}{0.5,0,0}
\definecolor{darkblue}{rgb}{0,0,0.5}
\definecolor{firebrick}{rgb}{0.75,0.125,0.125}
\definecolor{darkgreen}{rgb}{0,0.5,0}
\hypersetup{urlcolor=darkblue,
	    citecolor=darkgreen,
	    linkcolor=firebrick}

\usepackage{orcidlink}

\long\def\exclude#1{}

\newcommand{\ie}{{i.e.}}

\newcommand{\eg}{{e.g.}}
\newcommand{\Eg}{{E.g.}}

\newcommand{\eq}{Eq.}

\newcommand{\fig}{Fig.}

\newcommand{\Refe}{Ref.}
\newcommand{\Refes}{Refs.}
\newcommand{\equ}[1]{\eq~(\ref{equ:#1})}
\newcommand{\figu}[1]{\fig~\ref{fig:#1}}
\newcommand{\orcid}[1]{\href{https://orcid.org/#1}{\includegraphics[width=10pt]{orcid.pdf}}}
\bibpunct{[}{]}{,}{n}{}{,}

\begin{document}

\title{Searches for dark matter decay with ultra-high-energy neutrinos endure backgrounds}

\author{Damiano F.~G.~Fiorillo 
\orcidlink{0000-0003-4927-9850}} 
\email{damiano.fiorillo@nbi.ku.dk}
\affiliation{Niels Bohr International Academy, Niels Bohr Institute, University of Copenhagen, 2100 Copenhagen, Denmark}

\author{V\'ictor B.~Valera
\orcidlink{0000-0002-0532-5766}} 
\email{vvalera@nbi.ku.dk}
\affiliation{Niels Bohr International Academy, Niels Bohr Institute, University of Copenhagen, 2100 Copenhagen, Denmark}

\author{Mauricio Bustamante
\orcidlink{0000-0001-6923-0865}} 
\email{mbustamante@nbi.ku.dk}
\affiliation{Niels Bohr International Academy, Niels Bohr Institute, University of Copenhagen, 2100 Copenhagen, Denmark}

\author{Walter Winter
\orcidlink{0000-0001-7062-0289}} 
\email{walter.winter@desy.de}
\affiliation{Deutsches Elektronen-Synchrotron DESY, Platanenallee 6, 15738 Zeuthen, Germany}

\date{\today}

\begin{abstract}
 Next-generation ultra-high-energy (UHE) neutrino telescopes, presently under planning, will have the potential to probe the decay of heavy dark matter (DM) into UHE neutrinos, with energies in excess of $10^7$~GeV.  Yet, this potential may be deteriorated by the presence of an unknown background of UHE neutrinos, cosmogenic or from astrophysical sources, not of DM origin and seemingly large enough to obscure the DM signature.  We show that leveraging the angular and energy distributions of detected events safeguards future searches for DM decay against such backgrounds. We focus on the radio-detection of UHE neutrinos in the planned IceCube-Gen2 neutrino telescope, which we model in state-of-the-art detail. We report promising prospects for the discovery potential of DM decay into UHE neutrinos, the measurement of DM mass and lifetime, and limits on the DM lifetime, despite the presence of a large background, without prior knowledge of its size and shape.
\end{abstract}

\maketitle

\section{Introduction}
\label{sec:introduction}

About $85\%$ of the matter in the Universe is dark, not interacting electromagnetically nor strongly. Evidence for dark matter (DM) comes from velocity dispersion and rotation curves in galaxies and galaxy clusters~\cite{Oort:1932, Zwicky:1933gu, Zwicky:1937zza, Rubin:1985ze, Begeman:1991iy}, gravitational lensing measurements in galaxy clusters and collisions of galaxy clusters~\cite{1990ApJ...350...23B, Clowe:2006eq}, the cosmic microwave background (CMB) anisotropy, and the large-scale structure of the Universe~\cite{COBE:1992syq, WMAP:2010sfg, SDSS:2009ocz, DiMatteo:2007sq, Boylan-Kolchin:2009alo}. Because this evidence relates to the gravitational effect that DM has on the visible Universe, it provides little guidance to understand other possible interactions between DM and Standard-Model particles that could reveal its nature.

\begin{figure}[ht!]
 \includegraphics[width=\columnwidth]{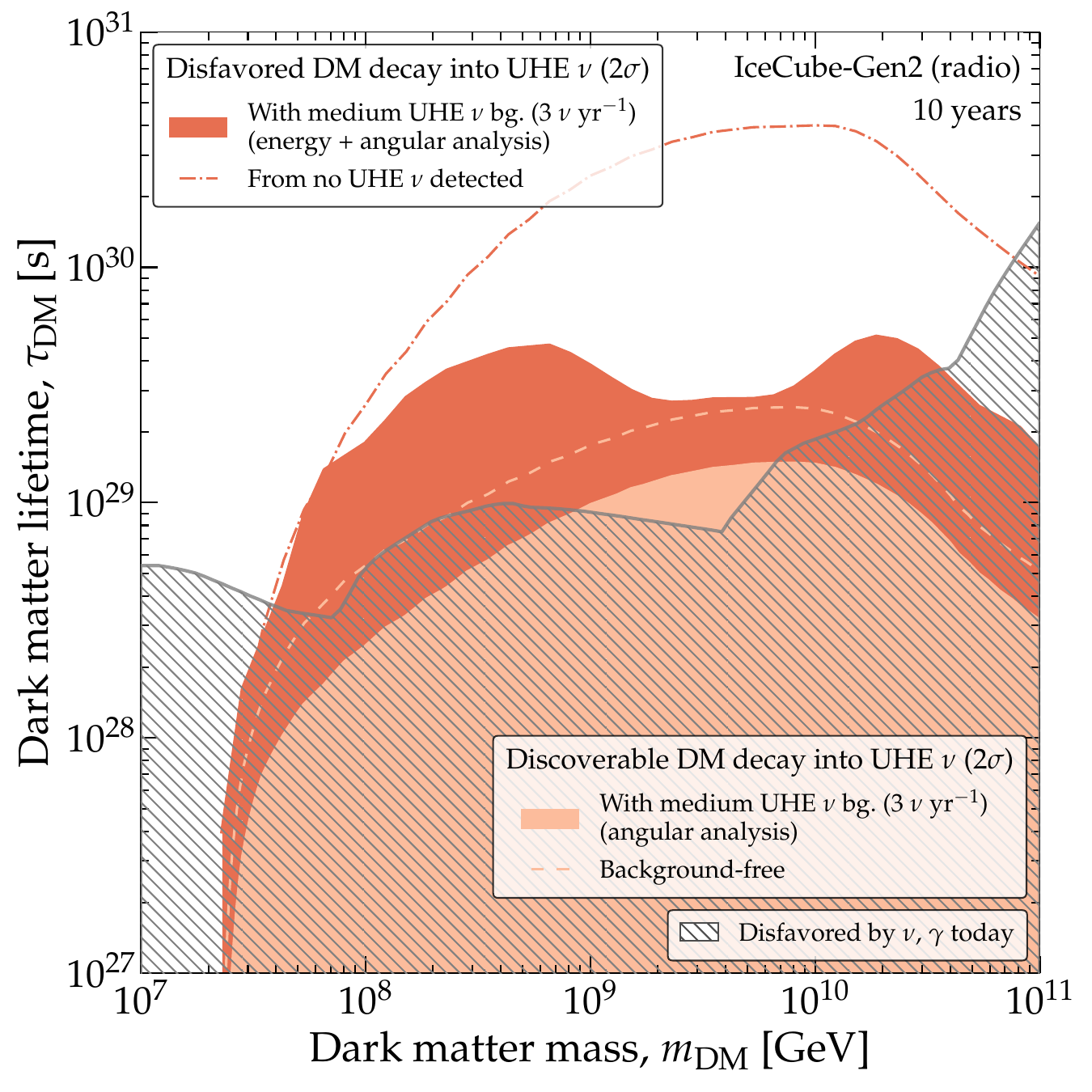}
 \caption{\textbf{\textit{Forecasts of discovery and constraints of dark matter decay into UHE neutrinos in the radio array of IceCube-Gen2.}}  {\it Dark matter (DM) decay can be discovered or constrained even in the presence of a medium-sized non-DM cosmogenic background flux of UHE neutrinos (\figu{diffuse_fluxes}).}  Forecasts are based on state-of-the-art projected samples of detected events with energies over $10^7$~GeV, use baseline choices of the detector angular and (logarithmic) energy resolution, $\sigma_{\Omega} = 3^\circ$ and $\sigma_{\epsilon} = 0.1$, and assume no knowledge of the non-DM neutrino background in the analysis of projected data.  For discovery, forecasts use the angular distribution of events; for constraints, their joint angular and energy distribution.  This figure assumes the Navarro-Frenk-White density profile for Galactic DM.  Existing lower limits on the DM lifetime are from the upper bound on the UHE neutrino flux from IceCube~\cite{IceCube:2018fhm}, KASCADE-Grande~\cite{KASCADEGrande:2017vwf}, the Telescope Array~\cite{TelescopeArray:2018rbt}, and the Pierre Auger Observatory~\cite{Veberic:2017hwu}; see \Refe~\cite{Arguelles:2022nbl}.  See Sec.~\ref{sec:discovery} for details on discovery (especially Figs.~\ref{fig:discovery_prospects}--\ref{fig:flux_reconstructed}) and Sec.~\ref{sec:bounds} for details on bounds (especially Figs.~\ref{fig:bounds_medium_bg} and \ref{fig:bounds}). \vspace*{-1cm} }\label{fig:summary}
\end{figure}

The absence of such guidance has spurred a broad program to understand the nature of DM, from theory and experiment.  From theory, diverse candidates have been proposed as DM constituents; see, \eg, \Refe~\cite{Bertone:2016nfn} for a historical review.  These include novel particles, such as weakly interacting massive particles~\cite{Kolb:1990vq, Griest:2000kj}, axions~\cite{Adams:2022pbo}, Majorons~\cite{Rothstein:1992rh, Berezinsky:1993fm, Brune:2018sab}, and sterile neutrinos~\cite{Boyarsky:2018tvu}, and non-particle candidates, such as primordial black holes~\cite{Carr:2021bzv}.  From experiment, searches for these candidates follow four complementary strategies: collider searches, which attempt to produce DM in high-energy particle collisions; direct DM searches, which look for Galactic DM scattering on dense detector targets; astrophysical searches, which look for the impact that DM would have on cosmic particles; and indirect DM searches, which look for products of DM self-annihilation or decay.  

We focus on indirect searches for the decay of heavy DM particles, with masses in excess of 10~PeV, into neutrinos.  Our choice is motivated by upcoming experimental capabilities (more on this later) that, for the first time, could allow us to probe DM decay using ultra-high-energy (UHE) neutrinos, with energies in excess of 10~PeV.  (We do not consider DM self-annihilation because, for heavy DM, its cross section is strongly constrained by unitarity bounds~\cite{Griest:1989wd, Smirnov:2019ngs, Chianese:2021htv}.)

Already in the last decade, the breadth of DM indirect searches widened after the discovery by the IceCube neutrino telescope of high-energy neutrinos of cosmic origin, with energies between $10$~TeV and $10$~PeV~\cite{IceCube:2013cdw, IceCube:2013low, IceCube:2014stg, IceCube:2015qii, IceCube:2016umi, Ahlers:2018fkn, IceCube:2020wum}.  In fact, initially the discovery led to speculation that heavy DM decaying to neutrinos could explain their flux in the 10--100~TeV range~\cite{Esmaili:2013gha, Feldstein:2013kka, Bai:2013nga, Ema:2013nda, Esmaili:2014rma, Bhattacharya:2014vwa, Murase:2015gea, Chianese:2016opp, Chianese:2016kpu, Chianese:2017nwe, Bhattacharya:2019ucd, Chianese:2019kyl}.  Nowadays, more conventional astrophysical explanations are favored~\cite{Murase:2012xs,Murase:2015xka,Capanema:2020rjj, Capanema:2020oet}, but IceCube observations still set competitive bounds on the DM lifetime and self-annihilation cross section for DM masses below $10$~PeV~\cite{IceCube:2022clp, Arguelles:2022nbl, IceCube:2023ies}. These bounds are complementary to the ones obtained from gamma-ray observations in a similar mass range; see, \eg, \Refes~\cite{Cohen:2016uyg,LHAASO:2022yxw,Arguelles:2022nbl}

In the next decade, a host of new neutrino telescopes, presently in different stages of planning, design, and prototyping, will target the long-sought discovery of UHE neutrinos, between $100$~PeV and $10$~EeV, that were first predicted in 1969~\cite{Berezinsky:1969erk}. They include AugerPrime~\cite{PierreAuger:2016qzd}, BEACON~\cite{Wissel:2020sec}, EUSO-SPB2~\cite{Adams:2017fjh}, GCOS~\cite{Horandel:2021prj}, GRAND~\cite{GRAND:2018iaj}, POEMMA~\cite{POEMMA:2020ykm},  PUEO~\cite{Deaconu:2019rdx}, RNO-G~\cite{RNO-G:2020rmc}, TAROGE~\cite{Nam:2020hng},  and the radio array of IceCube-Gen2, the envisioned upgrade of IceCube~\cite{IceCube-Gen2:2020qha}.  Ultra-high neutrinos will bring new insight into astrophysics~\cite{Ackermann:2019ows, Ackermann:2022rqc, AlvesBatista:2021eeu} and fundamental physics~\cite{Ackermann:2019cxh, Arguelles:2019rbn, Berryman:2022hds, MammenAbraham:2022xoc, Ackermann:2022rqc, AlvesBatista:2021eeu}. In particular, they will allow us to test the decay of heavier DM particles, with masses from $10$~PeV to $100$~EeV; see, \eg, \Refes~\cite{Guepin:2021ljb, Chianese:2021htv, Arguelles:2022nbl}.  

However, the capacity of UHE neutrino telescopes to probe DM decay critically depends on an unknown quantity: the diffuse flux of UHE neutrinos that do not originate from DM decay, and that acts as a background to DM searches.  If this background is large, it could  obscure more subtle signatures of neutrinos from DM decay. These background neutrinos---hereafter dubbed ``non-DM neutrinos''---are expected from the interaction of ultra-high-energy cosmic-ray (UHECRs) with ambient matter or radiation inside the extragalactic astrophysical sources where they are accelerated---\ie, {\it astrophysical} neutrinos---or with cosmic photon backgrounds during their propagation to Earth---\ie, {\it cosmogenic} neutrinos~\cite{Greisen:1966jv, Zatsepin:1966jv, Berezinsky:1969erk}.  We expand on them later (Sec~\ref{sec:fluxes_astro}).  

The situation is worsened by the large variety, in size and shape, in the current theoretical predictions of UHE astrophysical and cosmogenic neutrino fluxes; see, \eg, Fig.~2 in \Refe~\cite{Valera:2022wmu}.  Without a firm estimate of the non-DM UHE neutrino background, it would seem that the mere discovery of UHE neutrinos may be insufficient to establish whether they originate from DM decay or not.  Were the possibility of a background of non-DM UHE neutrinos ignored, the evidence for or against DM decay could be interpreted erroneously.

We show that these difficulties can be overcome by leveraging known differences between the distributions in energy and arrival directions of UHE neutrinos from DM decay and non-DM UHE neutrinos.  Regarding energy, neutrinos from DM decay are produced predominantly at an energy of $E_\nu\sim m_\mathrm{DM}/2$, where $m_\mathrm{DM}$ is the mass of the DM particle, whereas the spectrum of non-DM neutrinos is expected to be relatively extended in energy.  Regarding direction, the flux of neutrinos from DM decay should peak towards the Galactic Center (GC), where DM is concentrated, whereas the diffuse flux of non-DM neutrinos is expected to be isotropic.  The above features are essential and generic to UHE neutrinos of DM and non-DM origin alike, and are broadly present in models of their fluxes.  By relying on them, our methods apply broadly, regardless of the specific nature of the DM particle, of the relative size of the fluxes of neutrinos from DM decay and of non-DM neutrinos, and of the specific shape of the energy spectrum of non-DM neutrinos. 

Our strategy is similar to studies of the decay of TeV--PeV DM that use IceCube data (see, \eg, \Refe~\cite{IceCube:2022clp}), but has one important advantage.  In the TeV--PeV range, it is possible that non-DM astrophysical processes produce an excess of neutrinos towards the GC that could obfuscate a signal of neutrinos from DM decay~\cite{Ahlers:2013xia, Ahlers:2015moa, Neronov:2015osa, Carceller:2016upo, Neronov:2016bnp, Denton:2017csz, IceCube:2019scr, Kheirandish:2020upj, Vance:2021yky, Kovalev:2022izi, Sudoh:2022sdk, DelaTorreLuque:2022ats, Sudoh:2023qrz, IceCube:GC2023}.  In contrast, in the UHE range, we expect no astrophysical process to produce neutrinos towards the GC, making the search for DM decay cleaner.

We gear our forecasts to the radio-detection of UHE neutrinos in IceCube-Gen2, since it is among the largest upcoming neutrino telescopes under consideration and is presently in an advanced stage of planning.  We model neutrino detection via the same state-of-the-art simulations used in \Refes~\cite{Valera:2022ylt, Fiorillo:2022ijt, Valera:2022wmu}, which account for UHE neutrino propagation inside Earth, detector geometry, energy- and direction-dependent detector response, and energy and angular detector resolution.

Figure~\ref{fig:summary} summarizes our main findings; we defer details to later. They are two-fold: on the discovery of DM decay and on lower limits on the DM lifetime.  For the first time, we report robust discovery prospects for UHE neutrinos from DM decay, \ie, the values of DM mass and lifetime that would allow us not only to detect UHE neutrinos, but also to claim that they originate from DM decay, at least partially.  In \figu{summary}, the presence of about 30 neutrinos from DM decay in a 10-year event sample, would allow us to claim their DM origin. Separately, we find that while the background of non-DM UHE neutrinos weakens the lower limits on DM lifetime, an energy and angular analysis mitigates this weakening, keeping the limits competitive with present-day ones, at worst.

\textit{\textbf{The overarching message of our results is that, despite our ignorance of the background of astrophysical and cosmogenic UHE neutrinos, the discovery of UHE neutrinos will constitute a sensitive probe of heavy DM decay.}}  We present our results and methods to inform future forecasts and searches.

This paper is structured as follows.  In Sec.~\ref{sec:fluxes} we discuss the main features of DM and non-DM neutrino production, and highlight the models that we choose as benchmark for this work.  In Sec.~\ref{sec:detection} we describe how we compute UHE neutrino-induced event rates at IceCube-Gen2.  In Sec.~\ref{sec:discovery} we obtain the prospects of IceCube-Gen2 for the discovery of DM neutrinos.  In Sec.~\ref{sec:bounds} we forecast bounds on the DM lifetime if no evidence for DM decay is found.  In Sec.~\ref{sec:summary}, we conclude.


\section{Fluxes of UHE neutrinos}
\label{sec:fluxes}

\textit{The diffuse flux of UHE astrophysical and cosmogenic neutrinos, itself a target of discovery~\cite{Ackermann:2019ows, Valera:2022wmu, Ackermann:2022rqc, MammenAbraham:2022xoc}, could be a background to searches for the diffuse flux of UHE neutrinos from DM decay.  Fortunately, these fluxes differ in their distributions in energy and direction.  In energy, the non-DM background neutrino flux is spread out, while that of neutrinos from DM decay is more concentrated.  In direction, the non-DM background neutrino flux is isotropic, while that of neutrinos from DM decay peaks towards the Galactic Center.  We review these features below; later (Secs.~\ref{sec:discovery} and \ref{sec:bounds}), we use them to distinguish between the fluxes.}

\begin{figure*}
 \includegraphics[width=\textwidth]{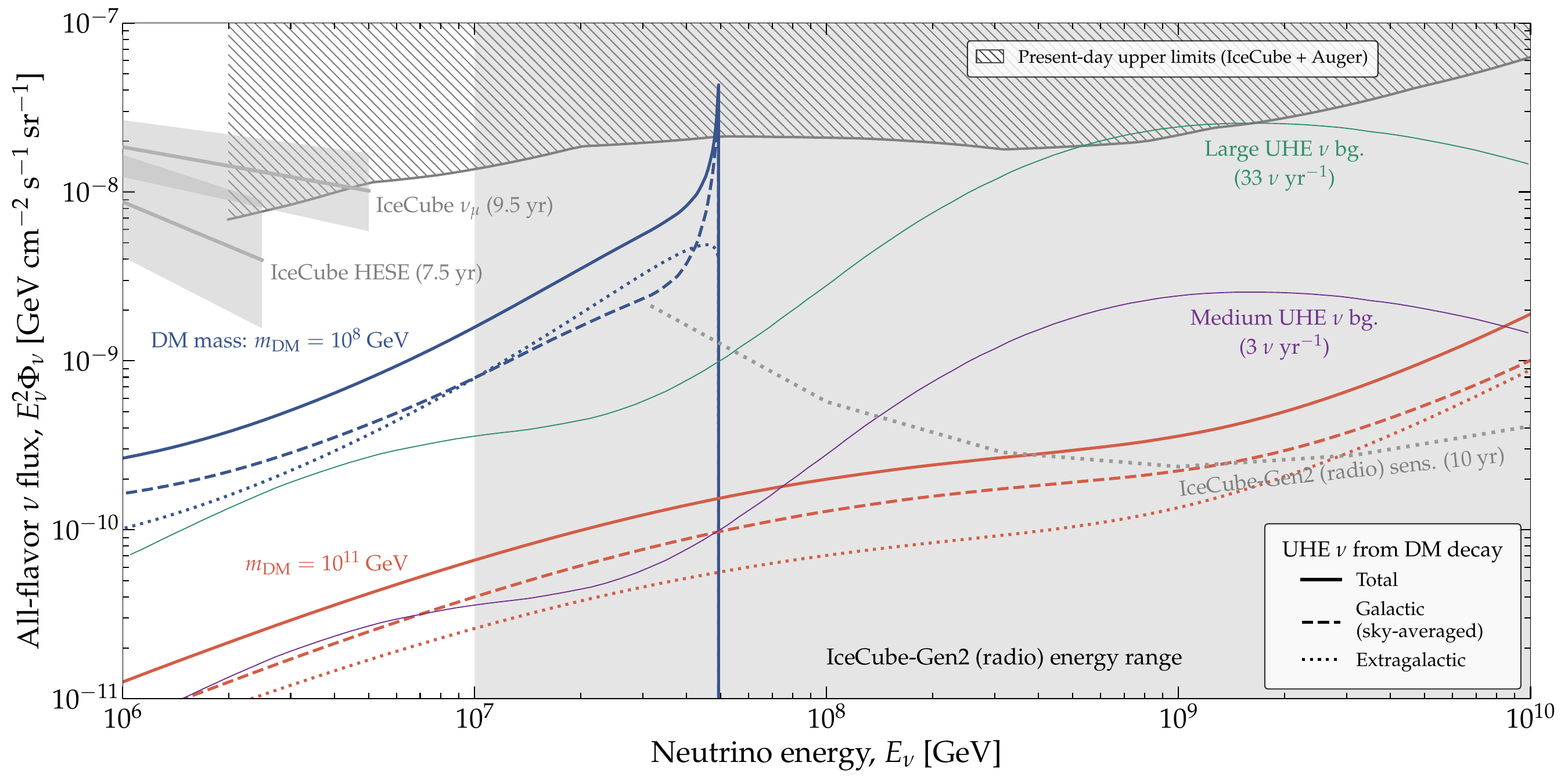}
 \caption{\textbf{\textit{Diffuse flux of UHE neutrinos from DM decay and non-DM background flux.}}  The flux from the decay of a DM particle, $\chi$, into neutrinos (\ie, $\chi \to \nu + \bar{\nu}$) is shown for illustrative choices of the DM mass, $m_{\rm DM} = 10^8$~GeV and $10^{11}$~GeV, and of the DM lifetime, $\tau = 10^{29}$~s, and computed following Sec.~\ref{sec:fluxes_dm}.  Later, we vary the values of the DM mass and lifetime in our forecasts.  The neutrino flux from DM decay is separated into its sky-averaged Galactic component and its isotropic extragalactic component.  (We do not show in this figure the primary-neutrino contribution to the Galactic component, since it is a singular monochromatic line at $E_\nu=m_\mathrm{DM}/2$, but we include it in our calculations; see Sec.~\ref{sec:fluxes_dm}.)  The two non-DM isotropic UHE neutrino fluxes are the benchmarks that we use in our work. The large benchmark flux is the cosmogenic neutrino flux predicted by Bergman \& van Vliet~\cite{Anker:2020lre} based on cosmic-ray data from the Telescope Array~\cite{Tsunesada:2017aaq, Bergman:2017ikd}.  The medium benchmark flux is 10\% of that.  The projected sensitivity of the radio array of IceCube-Gen2 is from \Refe~\cite{IceCube-Gen2:2021rkf}.  Present-day upper limits on the flux of UHE neutrinos are from IceCube~\cite{IceCube:2018fhm} and the Pierre Auger Observatory~\cite{PierreAuger:2019ens}.  For comparison, we show the tail end of the present-day IceCube measurements of TeV--PeV neutrinos~\cite{IceCube:2020wum, IceCube:2021uhz}.  See Sec.~\ref{sec:fluxes} for details.}\label{fig:diffuse_fluxes}
\end{figure*}


\subsection{UHE astrophysical and cosmogenic neutrinos}
\label{sec:fluxes_astro}

UHE neutrinos are expected from the interaction of UHECR protons, with energies $E_p \gtrsim 100$~PeV, with ambient matter~\cite{Margolis:1977wt, Stecker:1978ah, Kelner:2006tc} or radiation~\cite{Stecker:1978ah, Mucke:1999yb, Hummer:2010vx}, inside the extragalactic astrophysical sources where they are accelerated---\ie, {\it astrophysical} neutrinos---or with cosmic photon backgrounds during their propagation in extragalactic space---\ie, {\it cosmogenic} neutrinos.  These interactions produce high-energy pions, and other intermediate particles, that promptly decay into high-energy neutrinos via $\pi^- \to \mu^- + \bar{\nu}_\mu$, followed by $\mu^- \to e^- + \bar{\nu}_e + \nu_\mu$, and their charge-conjugated processes, where each neutrino has energy $E_\nu \simeq E_p/20$.  

We focus on the diffuse UHE neutrino flux, \ie, the sum of the UHE neutrino emission---astrophysical or cosmogenic---from all sources, across all redshifts.  The cosmogenic neutrino flux is isotropic, since extragalactic magnetic fields scramble the trajectories of neutrino-producing UHECRs; see \figu{sky_map_flux}.  The angular distribution of the astrophysical neutrino flux reflects that of the neutrino sources in the sky.  Because UHE neutrino sources are in all likelihood extragalactic, we assume that they are isotropically distributed, and so the diffuse neutrino flux from them is isotropic, too.  The discovery of the diffuse flux of astrophysical and cosmogenic neutrinos is one of the main goals of the next generation of neutrino telescopes~\cite{Ackermann:2019ows, Ackermann:2022rqc, Valera:2022ylt, Valera:2022wmu}.    (The associated discovery of point sources of UHE neutrinos is explored in \Refes~\cite{Fang:2016hyv, Fang:2016hop, Fiorillo:2022ijt}.)  Yet, in our work, they represent a background to the discovery of neutrinos from DM decay.

Cosmogenic neutrinos were first proposed in the late 1960s~\cite{Berezinsky:1969erk}, as a natural consequence of the interaction of UHECRs on the CMB~\cite{Greisen:1966jv, Zatsepin:1966jv}. They constitute a nearly guaranteed contribution in the UHE neutrino range, since their production only relies on the existence of UHECRs and of the CMB (and also of the extragalactic background light).  The flux of cosmogenic neutrinos depends on the properties of UHECRs---their spectrum, maximum energies, and mass composition---and of their sources---their distribution in redshift.  Because these properties are known uncertainly~\cite{AlvesBatista:2019tlv}, the flux predictions vary widely, in size and shape; see, \eg, \Refes~\cite{Aloisio:2009sj, Kotera:2010yn, Ahlers:2012rz, Fang:2013vla, Padovani:2015mba, Fang:2017zjf, Romero-Wolf:2017xqe, AlvesBatista:2018zui, Heinze:2019jou, Muzio:2019leu, Rodrigues:2020pli, Anker:2020lre, IceCube:2020wum, Muzio:2021zud, IceCube:2021uhz}.

Astrophysical UHE neutrinos are produced inside astrophysical sources. In this case, the target photons need not be the CMB, but low-energy photons present in the environments in which UHECRs are injected.  Flux predictions are made more complex because they depend also on the physical conditions inside the sources, including the shape of the photon spectra, the matter density, and the geometry of the neutrino production region.  Numerous models have been proposed for various candidate source classes, including active galactic nuclei (AGN)~\cite{Mannheim:1995mm, Atoyan:2001ey, Atoyan:2002gu, Alvarez-Muniz:2004xlu, Murase:2014foa, Kimura:2014jba, Murase:2014foa, Padovani:2015mba, Palladino:2018lov, Rodrigues:2020pli, Righi:2020ufi, Kimura:2020thg, Neronov:2020fww}, gamma-ray bursts (GRBs)~\cite{Paczynski:1994uv, Waxman:1997ti, Murase:2006mm,    Bustamante:2014oka, Senno:2015tsn, Pitik:2021xhb, Rudolph:2022dky, Rudolph:2022ppp}, newborn pulsars~\cite{Fang:2013vla, Fang:2014qva}, and tidal disruption events (TDEs)~\cite{Farrar:2008ex, Wang:2011ip, Dai:2016gtz, Senno:2016bso, Lunardini:2016xwi, Zhang:2017hom, Guepin:2017abw, Winter:2020ptf, Winter:2022fpf}. In some models, the diffuse astrophysical neutrino flux can be comparable or larger than the cosmogenic neutrino flux; \eg, \Refe~\cite{Rodrigues:2020pli}. 

Thus, there is a large number of competing theoretical predictions of the cosmogenic and astrophysical UHE neutrino flux; see Fig.~2 in \Refe~\cite{Valera:2022wmu} and Fig.~6 in \Refe~\cite{Valera:2022ylt} for an overview.  The range of predicted UHE neutrino fluxes spans several orders of magnitude.  The highest flux predictions~\cite{Anker:2020lre} would yield about 30 events per year in the radio array of IceCube-Gen2, making them easily discoverable; the lowest~\cite{Heinze:2019jou, Rodrigues:2020pli}, less than one event in 10 years, making them undiscoverable (see Fig.~1 and Table~I in \Refe~\cite{Valera:2022wmu} for details).  

Most flux predictions share some common features; \eg, they can be roughly described as a power-law flux---from neutrino production via proton-matter interactions---a bump-like flux---from neutrino production via proton-photon interactions---or a combination of both.  The resemblance between different flux predictions is largely superficial, since they differ in a number of important assumptions, \ie, the identity of the neutrino sources, the physical conditions in the region of neutrino production, the neutrino production mechanism, and the UHECR observations on which the neutrino predictions are based.  Regardless, in our forecasts below, we pivot on these superficial similarities and choose a benchmark background flux of UHE neutrinos that is representative of the range of theoretical predictions.

Figure~\ref{fig:diffuse_fluxes} shows the two  illustrative flux predictions that we select as benchmark ``non-DM'' UHE neutrino background for our analysis.  They represent a large background and an intermediate one; later, we complement them with a null-background scenario.  We base both on the cosmogenic neutrino flux predicted by Bergman \& van Vliet~\cite{Anker:2020lre} by fitting the simulated UHECR energy spectrum and mass composition at Earth to recent data from the Telescope Array (TA)~\cite{Tsunesada:2017aaq, Bergman:2017ikd}.  (This is flux model 4 in \Refes~\cite{Valera:2022ylt, Valera:2022wmu}.)  Because TA data favors a light UHECR mass composition and high maximum rigidity, the resulting cosmogenic neutrino flux is large: \figu{diffuse_fluxes} shows that it saturates the present-day upper limits from IceCube~\cite{IceCube:2018fhm} and the Pierre Auger Observatory~\cite{PierreAuger:2019ens}.  
\begin{description}
 \item[Large non-DM background] 
  This is the full cosmogenic neutrino flux predicted by Bergman \& van Vliet~\cite{Anker:2020lre}, which yields about 33 events per year in the radio array of IceCube-Gen2; see Figs.~3 and 4 and Table~I in \Refe~\cite{Valera:2022wmu}, and \figu{diff_event_rate} below.   Because this is as large a flux of UHE neutrinos as is allowed by present-day upper limits (\figu{diffuse_fluxes}), it is about the largest background of non-DM UHE neutrinos that we could face in a search for DM decay.
 \item[Intermediate non-DM background]
  This is the Bergman \& van Vliet cosmogenic flux scaled down to 10\% of its size, which yields about 3 events per year in the radio array of IceCube-Gen2.
 \item[Null background]
  The ideal scenario for the discovery of UHE neutrinos from DM is the absence of non-DM UHE neutrinos.  This has been the scenario adopted in previous forecasts of DM decay into UHE neutrinos~\cite{Guepin:2021ljb, Chianese:2021jke}.  We maintain it here as a baseline against which we compare our forecasts including a non-DM background.
\end{description}
Figure~\ref{fig:diffuse_fluxes} shows the all-flavor background flux, but when computing event rates (Sec.~\ref{sec:detection}) we sum the individual contributions of the fluxes of $\nu_e$, $\bar{\nu}_e$, $\nu_\mu$, $\bar{\nu}_\mu$, $\nu_\tau$,  and $\bar{\nu}_\tau$ for this model, as shown in Fig.~6 in \Refe~\cite{Valera:2022ylt}.  

For the purpose of discovering neutrinos from DM decay, what matters is not whether the non-DM UHE neutrino background is cosmogenic or astrophysical, but rather that its angular and energy distributions are different from those of the flux of neutrinos from DM decay (Sec.~\ref{sec:introduction}).  We point out these differences explicitly in Sec.~\ref{sec:fluxes_astro_vs_dm}.  Admittedly, in choosing a benchmark non-DM UHE neutrino background, we make a specific choice of the shape of its energy spectrum.  This choice is necessary to be able to generate simulated samples of detected events (Sec.~\ref{sec:detection}).  However, when analyzing these samples (Secs.~\ref{sec:bounds} and \ref{sec:discovery}), we do not assume knowledge of the size or shape of the non-DM background, but instead let them vary, just as an analysis of real detected data would.


\subsection{UHE neutrinos from dark matter decay}
\label{sec:fluxes_dm}

\begin{figure}
 \includegraphics[width=\columnwidth]{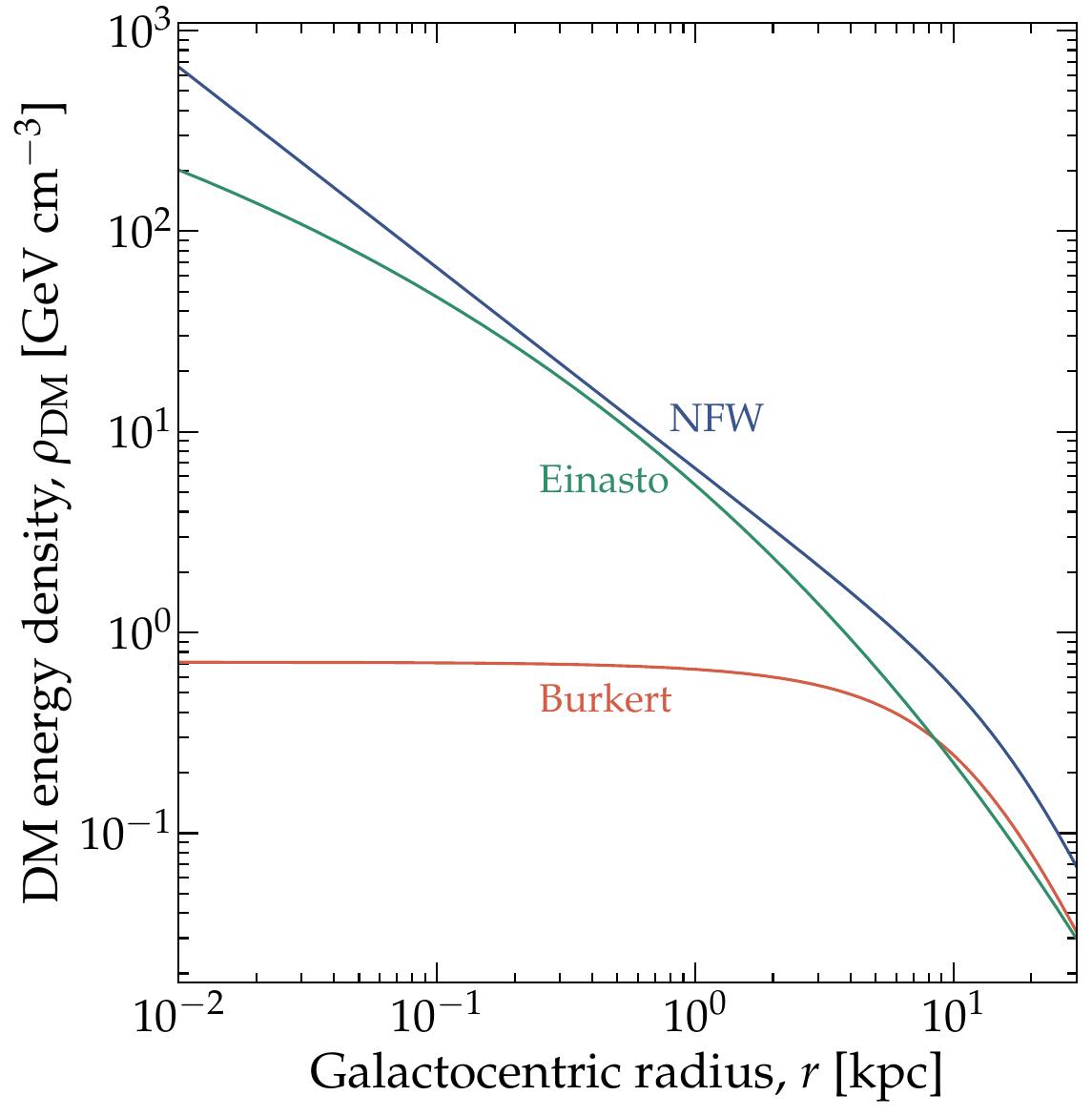}
 \caption{\textbf{\textit{Profiles of Galactic dark matter density.}}  For the Navarro-Frenk-White (NFW) profile~\cite{Navarro:1995iw}, we use \equ{dm_density_nfw}.  For the Einasto~\cite{Einasto:1965czb} and Burkert~\cite{Burkert:1995yz} profiles, we use the parametrizations and parameter choices from \Refe~\cite{Cirelli:2010xx}; see \equ{dm_density_burkert}.  Our main results for DM discovery (Figs.~\ref{fig:discovery_prospects}, \ref{fig:reconstructed}, \ref{fig:flux_reconstructed}, and \ref{fig:discovery_prospects_3sigma}) and bounds (Figs.~\ref{fig:bounds_medium_bg}, \ref{fig:bounds}, and \ref{fig:bounds_large_bg}) are generated assuming the NFW profile.  Results generated assuming the Einasto profile (not shown) are similar. Figures~\ref{fig:discovery_prospects_nfw_vs_burkert} and \ref{fig:bounds_nfw_vs_burkert} contrast results obtained assuming the NFW and Burkert profiles.}
 \label{fig:dm_profiles}
\end{figure}

\begin{figure}
 \includegraphics[width=\columnwidth]{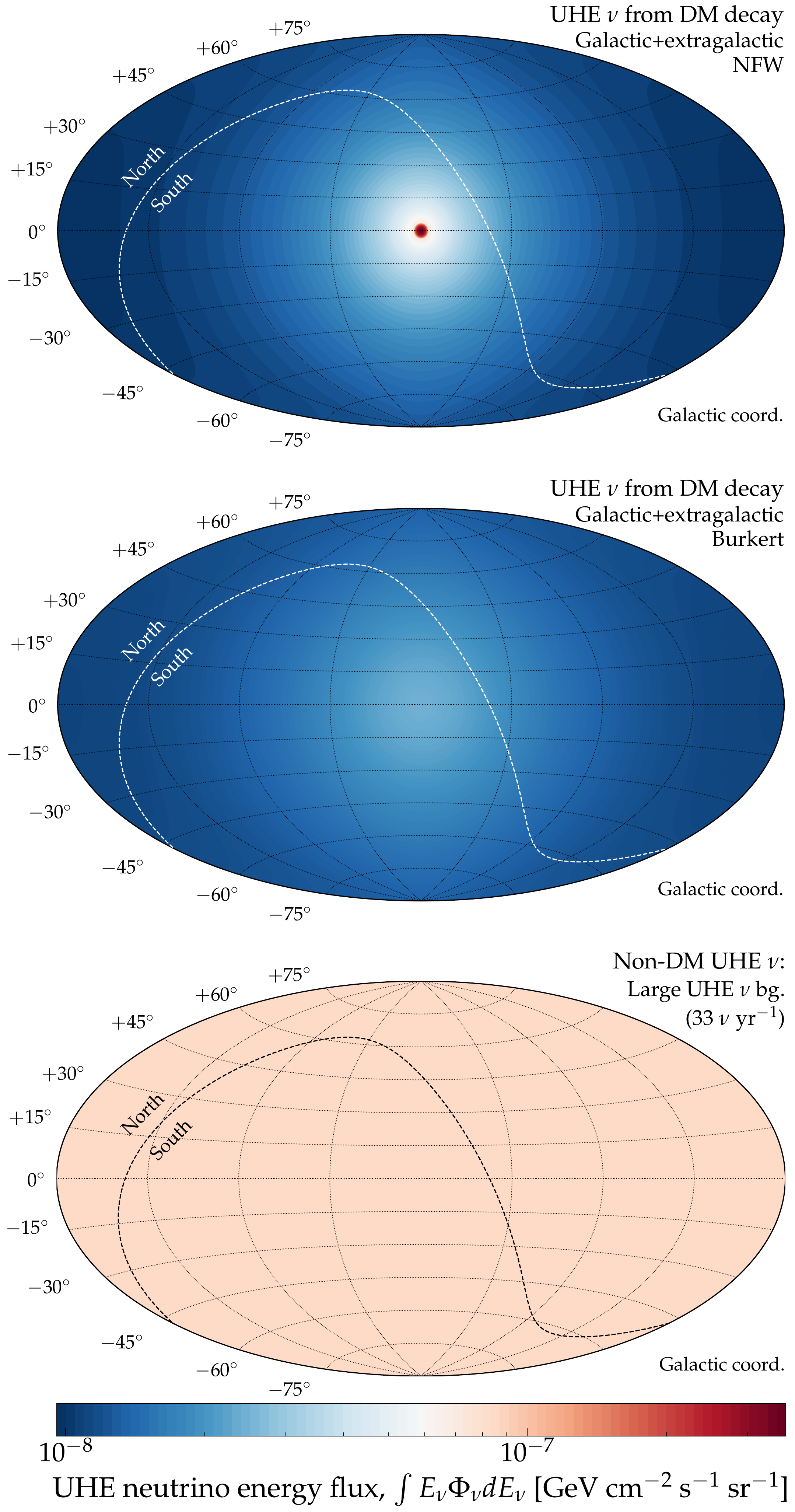}
 \caption{\textbf{\textit{Angular distribution of the diffuse flux of UHE neutrinos.}} {\it Top:} Neutrinos from DM decay, for an NFW profile.  We show the distribution computed using illustrative values of DM mass and lifetime, $m_\mathrm{DM}=10^8$~GeV and $\tau_\mathrm{DM}=10^{29}$~s, but for other values the angular distribution is very similar. {\it Center:} Same for a Burkert profile. {\it Bottom:} Neutrinos from the non-DM isotropic background of UHE neutrinos.  The flux is isotropic, so the sky map is shaded uniformly.  We show the distribution from our large benchmark background, \ie, the cosmogenic neutrino flux by Bergman \& van Vliet~\cite{Anker:2020lre}, but for other choices of the isotropic background flux the angular distribution is very similar. \vspace*{-0.8cm}}\label{fig:sky_map_flux}
\end{figure}

The decay of a heavy DM particle, $\chi$, with mass $m_{\rm DM} \gtrsim 10^7$~GeV, into Standard Model particles, leads to the production of high-energy neutrinos. The yield of neutrinos depends on the channels by which the DM particle decays. If the DM particle decays primarily into neutrinos, \ie, $\chi \to \bar{\nu} \nu$, the resulting neutrino flux has a primary contribution that is monoenergetic at $E_\nu = m_\mathrm{DM}/2$. We neglect the spread of the energy spectrum due to the thermal velocity of DM, since it is small~\cite{Hogan:2000bv, Bode:2000gq}.  A secondary contribution comes from electroweak corrections, generated from the emission, by the decay products, of off-shell $W$ and $Z$ bosons that promptly decay into neutrinos; this contribution is present even if DM does not primarily decay to neutrinos. In our analysis, we consider the neutrino flux made up of both primary and secondary contributions.

The electroweak corrections unavoidably give rise to gamma rays, electrons, positrons, protons, and anti-protons that are also amenable to indirect detection. Upper limits on their flux indirectly constrain the associated neutrino flux.  Notably, for most decay channels (\eg, for hadronic decay channels such as $\chi \to \overline{b}b$), the present-day upper limits on the gamma-ray flux are so strong that the projected limits on the associated UHE neutrino flux are comparable or weaker; see, \eg, \Refe~\cite{Chianese:2021jke} (see also \Refe~\cite{Das:2023wtk} for a comparison of the bounds from gamma-ray and cosmic-ray measurements). However, for leptonic (\eg, $\chi \to \overline{\tau}\tau$) and  neutrinophilic (\ie, $\chi \to \overline{\nu}\nu$) decay channels, in the DM mass range $m_{\rm DM} = 10^7$--$10^{10}$~GeV, the bounds from UHE neutrinos may be comparable or stronger than the present-day bounds from gamma rays. For this reason, we focus exclusively on the neutrinophilic decay channel, $\chi \to \overline{\nu}\nu$. We assume equal branching ratios for the decay into each of the three flavors, $\chi \to \overline{\nu}_e\nu_e$, $\chi \to \overline{\nu}_\mu\nu_\mu$, and $\chi\to\overline{\nu}_\tau\nu_\tau$.  We treat the flux of $\nu_e$, $\bar{\nu}_e$, $\nu_\mu$, $\bar{\nu}_\mu$, $\nu_\tau$, and $\bar{\nu}_\tau$ separately; later, we propagate each through the Earth (Sec.~\ref{sec:detection_propagation}) and compute its contribution to the detected event rate (Sec.~\ref{sec:detection_simulation}).

We compute the neutrino spectra numerically, using the public code \texttt{HDMSpectra}~\cite{Bauer:2020jay}, which evolves the particle showers initiated by DM decay, including in detail electroweak corrections, and yields the final-state products of shower evolution. The neutrino spectra from DM decay are notoriously hard to compute precisely, since the emission of soft collinear $W^\pm$ bosons leads to logarithmically enhanced terms $\propto \log^2(m_\mathrm{DM}/m_W)$, where $m_W$ is the mass of the $W$ boson, that need to be resummed~\cite{Ciafaloni:2010ti}; \texttt{HDMSpectra} accounts for this.
Thus, from \texttt{HDMSpectra} we obtain $dN_{\nu_\alpha}/dE_\nu$ and $dN_{\bar{\nu}_\alpha}/dE_\nu$ ($\alpha = e,\mu,\tau$), the number of $\nu_\alpha$ and $\bar{\nu}_\alpha$ emitted in a single DM decay per unit energy.  These spectra include also the primary monoenergetic contributions at $E_\nu = m_\mathrm{DM}/2$ .

The diffuse flux of neutrinos that reach the Earth is due to DM decays that occur inside the Galaxy (Gal) and in extragalactic space (EG), \ie,
\begin{equation}
 \frac{d\Phi_{\nu_\alpha}}{dE_\nu d\Omega_\nu}
 =
 \frac{d\Phi^\mathrm{Gal}_{\nu_\alpha}}{dE_\nu d\Omega_\nu}
 +
 \frac{d\Phi^\mathrm{EG}_{\nu_\alpha}}{dE_\nu d\Omega_\nu}
 \;,
\end{equation}
where $\Omega_\nu$ is the solid angle.  To compute the Galactic contribution, we integrate the neutrino spectrum from a single DM decay over the spatial distribution of DM in the Milky Way.  This makes the Galactic neutrino flux anisotropic, since it traces the density of Galactic DM.  To compute the extragalactic contribution, we integrate the neutrino spectrum from a single DM decay over the cosmological distribution of DM.  This makes the extragalactic neutrino flux isotropic.

The Galactic contribution of $\nu_\alpha$ is 
\begin{equation}
 \frac{d\Phi^\mathrm{Gal}_{\nu_\alpha}}{dE_\nu d\Omega_\nu}
 =
 \frac{dN_{\nu_\alpha}}{dE_\nu}\int_0^{\infty} \frac{\rho_\mathrm{DM}(s,b,l)}{4\pi \tau_\mathrm{DM} m_\mathrm{DM}}
 ds \;,
\end{equation}
where $s$ is the distance measured from the Earth, $b$ and $l$ are Galactic latitude and longitude and parametrize the neutrino incoming direction, $\tau_\mathrm{DM}$ is the DM lifetime, and $\rho_\mathrm{DM}$ is the density profile of DM in the Galaxy.

Figure~\ref{fig:dm_profiles} shows competing models of the Galactic DM density profiles.  The ``cuspy'' Navarro-Frenk-White (NFW)~\cite{Navarro:1995iw} profile---obtained from a numerical fit to $N$-body simulations of structure formation---and Einasto~\cite{Einasto:1965czb} profile---originally proposed to describe stellar systems and later extended to fit the DM halo---peak towards the GC.  The ``puffy'' Burkert profile~\cite{Burkert:1995yz}---obtained by a fit to the DM distribution in dwarf galaxies---instead plateaus to a core towards the GC.  We pick the NFW and Burkert profiles as representative of the two extremes of the ``cusp {\it vs.}~core'' uncertainty in Galactic DM profiles.   To produce our main results, in Figs.~\ref{fig:discovery_prospects}, \ref{fig:reconstructed}, \ref{fig:flux_reconstructed}, \ref{fig:discovery_prospects_3sigma}, \ref{fig:bounds_medium_bg}, \ref{fig:bounds}, and \ref{fig:bounds_large_bg}, we adopt the NFW profile; in Figs.~\ref{fig:bounds_medium_bg}, \ref{fig:bounds}, and \ref{fig:bounds_large_bg}, we contrast them against results obtained assuming the Burkert profile.  For the NFW profile, we use~\cite{Cirelli:2010xx}
\begin{equation}
 \label{equ:dm_density_nfw}
 \rho_\mathrm{DM}^{\rm NFW}(s,b,l)
 =
 \frac{\rho_0}{\left(\frac{r(s,b,l)}{r_c}\right)\left(1+\frac{r(s,b,l)^2}{r_c^2}\right)} \;,
\end{equation}
where $\rho_0 = 0.33$~GeV~cm$^{-3}$, $r_c = 20$~kpc, and
$r(s,b,l)=\sqrt{s^2+R_s^2-2 s R_s \cos b \cos l}$ is the Galactocentric radius, with $R_s = 8.5$~kpc. For the Burkert profile, we use
\begin{equation}
 \label{equ:dm_density_burkert}
 \rho^\mathrm{Burkert}_\mathrm{DM}(s,b,l
 )
 =\frac{\rho_s}{\left(1+\frac{r(s,b,l)}{r_s}\right)\left(1+\frac{r(s,b,l)^2}{r_s^2}\right)} \;,
\end{equation}
with $\rho_s=0.712$~GeV~cm$^{-3}$ and $r_s=12.67$~kpc~\cite{Cirelli:2010xx}.

The extragalactic contribution of $\nu_\alpha$ is
\begin{equation}
 \label{equ:nu_flux_eg}
 \frac{d\Phi^\mathrm{EG}_{\nu_\alpha}}{dE_\nu d\Omega_\nu}
 =
 \frac{\Omega_\mathrm{DM}\rho_c}{4\pi \tau_\mathrm{DM}m_\mathrm{DM}}\int_0^\infty \frac{dz}{H(z)}\left. 
 \frac{dN_{\nu_\alpha}}{dE_\nu}
 \right\vert_{E_\nu(1+z)} \;,
\end{equation}
where $z$ is the redshift, $\rho_c = 4.79\times 10^{-6}$~GeV~cm$^{-3}$ is the critical density of the Universe, $\Omega_\mathrm{DM}=0.265$ is the fraction of energy density of the Universe in the form of DM, $H(z) = H_0\sqrt{\Omega_\Lambda+\Omega_m (1+z)^3}$ is the Hubble parameter, $H_0 = 1.08\times 10^{-28}$~cm$^{-1}$~$h$ is the Hubble constant, with $h=0.674$, $\Omega_\Lambda = 0.685$ is the vacuum energy density, and $\Omega_m = 0.315$ is the matter energy density.  The right-hand side of \equ{nu_flux_eg} is evaluated at an energy $E_\nu (1+z)$ to compensate for the cosmological expansion.

Figure~\ref{fig:diffuse_fluxes} shows the resulting diffuse energy spectrum of UHE neutrinos from DM decay, integrated over all sky directions, for two benchmark values of the DM mass, and separated into the Galactic and extragalactic components only for illustration. The main features of the energy spectrum are a spike of neutrinos close to the  energy $E_\nu = m_\mathrm{DM}/2$, and a power-law tail at lower energies, from electroweak corrections and, in the case of the extragalactic component, from redshifting. The dominant component of the flux is the Galactic one, due to the nearby DM overdensity in the GC.  However, at energies close to the spike, a pile-up of neutrinos from the direct decay $\chi \to \overline{\nu}\nu$, redshifted to lower energies, causes the extragalactic contribution to dominate instead in a narrow energy range. Because we assume DM decay into neutrinos of all flavors with equal branching ratios (see above), the all-flavor flux in \figu{diffuse_fluxes} is split evenly among the three flavors and among neutrinos and anti-neutrinos.


\subsection{Non-DM neutrinos {\it vs.}~neutrinos from DM decay}
\label{sec:fluxes_astro_vs_dm}

The essential differences between the background flux of UHE non-DM neutrinos and the flux of UHE neutrinos from DM decay are in their energy spectrum and in their angular distribution in the sky.

\begin{description}
 \item[Energy spectrum]
  Figure~\ref{fig:diffuse_fluxes} shows that the energy spectrum of our benchmark non-DM background flux---which is typical of many flux predictions---is more spread out around its maximum compared to the energy spectrum of neutrinos from DM decay, which peaks sharply at $E_\nu = m_{\rm DM}/2$.  Yet, because of the spread of the latter towards lower energies due to redshifting (compounded, later, by the limited energy resolution of the detector), the differences in spectral shape are not as marked.  While the lack of a bump-like feature in the observed spectrum of UHE neutrinos would disfavor DM decay, its observation could be attributed either to DM decay or to a non-DM background flux.  Accordingly, the energy spectrum is not the driving factor to discover DM decay (Sec.~\ref{sec:discovery}), but supplements angular information to set constraints (Sec.~\ref{sec:bounds}).
 \item[Angular distribution]
  Figure~\ref{fig:sky_map_flux} shows the angular distribution of the total diffuse flux of UHE neutrinos, integrated over energy.  This reveals the critical difference between non-DM and DM neutrinos: the flux of non-DM neutrinos is isotropic, while the flux of DM neutrinos peaks towards the GC, where, under the NFW profile, it is about a factor $20$~larger than in the rest of the sky.  This contrast is the main driving factor to discover DM decay and place constraints on it.  For the puffier Burkert profile, the contrast is milder, which weakens both prospects, as we show later.  In either case, because the GC is in the Southern Hemisphere, neutrinos from DM decay coming from this direction are not attenuated by their passage through Earth before reaching IceCube-Gen2 (Sec.~\ref{sec:detection_propagation}), making it particularly sensitive to this signal. 
\end{description}

Below (Sec.~\ref{sec:detection}), we show that the above differences between the fluxes are mirrored, albeit imperfectly, by corresponding differences in the energy and angular distributions of detected events.


\section{Detection of UHE neutrinos}
\label{sec:detection}

{\it To make realistic forecasts of probes of DM decay into UHE neutrinos, we compute in detail their propagation inside the Earth and their radio-detection in our detector of choice, the radio array of IceCube-Gen2.}


\subsection{Neutrino propagation inside the Earth}
\label{sec:detection_propagation}

Upon reaching Earth, UHE neutrinos propagate from its surface, through its interior, to the detector, IceCube-Gen2, situated at the South Pole.  While propagating, neutrinos interact with matter underground.  Because the neutrino-nucleon cross section, $\sigma_{\nu N}$, grows with neutrino energy (at ultra-high energies, roughly as $\sigma_{\nu N} \propto E_\nu^{0.363}$~\cite{Gandhi:1998ri}), these interactions appreciably attenuate the flux of neutrinos that reaches the detector.  Roughly, neutrino interactions  attenuate the flux via an exponential dampening factor $e^{-\sigma_{\nu N} L}$, where $L$ is the distance traveled underground.  Thus, the attenuation grows with neutrino energy and with distance traveled.  

At ultra-high energies, neutrinos interact with nucleons predominantly via deep inelastic scattering (DIS)~\cite{CTEQ:1993hwr, Conrad:1997ne, Formaggio:2012cpf, IceCube:2017roe, Bustamante:2017xuy, IceCube:2018pgc, IceCube:2020rnc}.  In it, the incoming neutrino scatters off of a parton---a quark or a gluon---of a nucleon are rest, $N$---a proton or a neutron ($N = p, n$).  The interaction is neutral-current (NC) if mediated by a $Z$ boson, \ie, $\nu_\alpha + N \to \nu_\alpha + X$ ($\alpha = e, \mu, \tau$), where $X$ represents final-state hadrons, or charged-current (CC) if mediated by a $W$ boson, \ie, $\nu_\alpha + N \to \alpha + X$.  The NC neutrino-nucleon cross section, $\sigma_{\nu_\alpha N}^{\rm NC}$, is about $1/3$ of the CC cross section, $\sigma_{\nu_\alpha N}^{\rm CC}$.  At these energies, the cross sections on proton and on neutron are very similar, and the cross sections for different neutrino flavors are nearly equal.  When propagating neutrinos inside the Earth, and also when computing detected event rates (Sec.~\ref{sec:detection_rates}), we treat separately the NC and CC interactions of neutrinos of different flavor, each with its own cross section.

In a DIS interaction, the final-state hadrons receive a fraction $y$---the inelasticity---of the energy of the interacting neutrino.  The final-state leptons receive the remaining fraction, $(1-y)$.  The inelasticity follows a probability distribution given by the differential cross sections, either $d\sigma_{\nu_\alpha N}^{\rm NC}/dy$ or $d\sigma_{\nu_\alpha N}^{\rm CC}/dy$.  The distributions peak at $y = 0$, but they are broad, and depend on the neutrino energy; see, \eg, Fig.~4 in \Refe~\cite{Valera:2022ylt}.  

Thus, as neutrinos propagate inside the Earth, NC interactions shift the neutrino flux to lower energies, while CC interactions deplete the flux.  (For $\nu_\tau$, the consecutive CC neutrino interactions and decays of the ensuing tauons---known as ``$\nu_\tau$ regeneration''---appreciably counteract the flux dampening; see, \eg, Fig.~8 in \Refe~\cite{Valera:2022ylt}.)  At ultra-high energies, the flux of upgoing neutrinos, with $\theta_z > 90^\circ$, where $\theta_z$ is the zenith angle measured from the South Pole, is nearly fully attenuated by the time it reaches the detector.  On the contrary, the flux of downgoing ($\theta_z < 90^\circ$) and horizontal ($\theta_z \approx 90^\circ$) neutrinos is attenuated appreciably, but is not completely depleted.  For illustration, see, \eg, Fig.~A2 in \Refe~\cite{Bustamante:2017xuy}, \Refe~\cite{Garcia:2020jwr}, and Figs.~10 \& 11 in \Refe~\cite{Valera:2022ylt}.  This makes the detection of UHE neutrinos more likely from these directions, provided there is sufficient detector response, which is the case for our modeling of IceCube-Gen2; we elaborate on this in Secs.~\ref{sec:detection_simulation} and \ref{sec:detection_rates}. 

We compute the propagation of UHE neutrinos inside the Earth as in \Refes~\cite{Valera:2022wmu, Fiorillo:2022ijt, Valera:2022ylt}, using the sophisticated propagation code \textsc{NuPropEarth} \cite{Garcia:2020jwr, NuPropEarth}.  It uses the recent BGR18 neutrino-nucleon DIS cross sections~\cite{Bertone:2016nfn}, the same ones that we use in Sec.~\ref{sec:detection_rates} to compute the rate of detected events.  \textsc{NuPropEarth} also accounts for $\nu_\tau$ regeneration, for energy losses of intermediate leptons during propagation, and for subleading neutrino interactions that, taken together, increase the flux attenuation by approximately an extra 10\%.  For the density profile of matter inside the Earth, we adopt the Preliminary Reference Earth Model~\cite{Dziewonski:1981xy}, with an added layer of surface ice 3~km thick to represent Antarctica, and account also for the radial change in the chemical composition of underground matter~\cite{Garcia:2020jwr}.  Finally, we model the volume of the neutrino detector---the radio array of IceCube-Gen2---as a cylinder of radius 12.6~km and height 1.50~km, buried vertically 100~m underground at the South Pole; see Fig.~7 in \Refe~\cite{Valera:2022ylt}.

In summary, given a flux of neutrinos at the surface of the Earth, from DM decay or from the non-DM background neutrino flux, we propagate it across many different directions to the detector.  We propagate separately the fluxes of $\nu_e$, $\bar{\nu}_e$, $\nu_\mu$, $\bar{\nu}_\mu$, $\nu_\tau$, and $\bar{\nu}_\tau$;. Below, we use their fluxes at the detector, $\Phi_{\nu_\alpha}^{\rm det}$ and $\Phi_{\bar{\nu}_\alpha}^{\rm det}$, to compute neutrino-induced event rates.


\subsection{UHE neutrino radio-detection at IceCube-Gen2}
\label{sec:detection_simulation}

Reference~\cite{Zas:1991jv} first proposed using the radio emission from UHE particles as a means to detect them.  Upon reaching the detector volume, an UHE neutrino may scatter off a nucleon in ice and produce a shower of high-energy particles.  As the shower travels, it accumulates an excess of electrons in its front that, after reaching shower maximum, is emitted as an impulsive coherent radio pulse, known as Askaryan radiation~\cite{Askaryan:1961pfb}.  For details, see \Refe~\cite{Schroder:2016hrv, Barwick:2022vqt}.  Because radio travels in ice subject only to mild attenuation, it may be detected using a sparse underground array of radio antennas, which makes it feasible to instrument a large volume that makes up for the potentially tiny fluxes of incoming UHE neutrinos.  

This is the strategy adopted by the planned radio array of IceCube-Gen2~\cite{IceCube-Gen2:2020qha}.  Of the proposed UHE neutrino telescopes~\cite{MammenAbraham:2022xoc, Ackermann:2022rqc, Guepin:2022qpl}, the radio array of IceCube-Gen2 is among the largest and in an advanced stage of planning~\cite{IceCube-Gen2:2020qha}.  Thus, we gear our forecasts to it.  However, our methods can be readily adapted to other upcoming UHE neutrino telescopes~\cite{MammenAbraham:2022xoc, Ackermann:2022rqc, Guepin:2022qpl}.  

To compute realistic projected event rates at the radio array of IceCube-Gen2, we follow the detailed procedure introduced in \Refe~\cite{Valera:2022ylt}, which uses an estimated detector response based on state-of-the-art simulations.  This has been used already to forecast the measurement of the UHE neutrino-nucleon cross section~\cite{Valera:2022ylt}, the discovery of UHE neutrino point sources~\cite{Fiorillo:2022ijt}, and the discovery of the diffuse flux of UHE neutrinos~\cite{Valera:2022wmu}.  Below, we only sketch the procedure and introduce necessary modifications to it; we defer to \Refe~\cite{Valera:2022ylt} for details. 

Upon reaching the detector volume, after propagating through the Earth (Sec.~\ref{sec:detection_propagation}), an UHE $\nu_\alpha$ of energy $E_\nu$ interacts with a nucleon at rest, $N$, typically via DIS (see above).  The ensuing particle shower has an energy $E_{\rm sh}$, a fraction of the parent neutrino energy.  For showers initiated by the NC DIS of $\nu_\alpha$ or $\bar{\nu}_\alpha$ of any flavor, only the final-state hadrons radiate~\cite{Garcia-Fernandez:2020dhb}, so $E_{\rm sh} = y E_\nu$.  For showers initiated by the CC DIS of a $\nu_e$ or $\bar{\nu}_e$, both the final-state electron and hadrons radiate, so $E_{\rm sh} = E_\nu$.  For showers initiated by the CC DIS of $\nu_\mu$, $\bar{\nu}_\mu$, $\nu_\tau$, or $\bar{\nu}_\tau$, only the final-state hadrons radiate, so $E_{\rm sh} = y E_\nu$.  As during propagation, at detection the value of the inelasticity follows $d\sigma_{\nu_\alpha}^{\rm NC}/dy$ and $d\sigma_{\nu_\alpha}^{\rm CC}/dy$, for which we adopt the BGR18~\cite{Bertone:2016nfn} calculation; see Fig.~4 in \Refe~\cite{Valera:2022ylt}.

The detector response is represented by its effective volume, which we treat separately for NC and CC showers, $V_{{\rm eff}, \nu_\alpha}^{\rm NC}$ and $V_{{\rm eff}, \nu_\alpha}^{\rm CC}$.  The effective volume depends on the shower energy and on the direction of the incoming neutrino.  It is generated by simulating the interaction of neutrinos in the detector volume, followed by the generation of Askaryan radiation, its propagation in ice, including changes in the index of refraction of ice with depth, and its detection in the two types of radio antennas envisioned in the array.  For the simulations we use \textsc{NuRadioReco}~\cite{Glaser:2019cws} and \textsc{NuRadioMC}~\cite{Glaser:2019rxw}, the same tools used by the IceCube-Gen2 Collaboration.  We adopt the same array design consisting of a combination of shallow and deep radio stations as in \Refe~\cite{Valera:2022ylt}.  The effective volume is least sensitive around $10^7$~GeV, grows with shower energy, and is relatively less sensitive for downgoing neutrinos ($\cos \theta_z \approx 1$); see Fig.~13 in \Refe~\cite{Valera:2022ylt}.  (Unlike common practice, the detector volume does not contain the effect of the attenuation of the neutrino flux underground.  This is contained separately, in $\Phi_{\nu_\alpha}^{\rm det}$.)

The differential event rate is obtained by convolving the neutrino flux that reaches the detector (Sec.~\ref{sec:detection_propagation}), $\Phi^{\rm det}_{\nu_\alpha}$, the effective volume, and the neutrino-nucleon cross section.  For $\nu_\alpha$, after an exposure time $T$, this is
\begin{widetext}
 \begin{eqnarray}
  \label{equ:spectrum_true}
  \frac{d^2N_{\nu_\alpha}}{dE_{\rm sh} d\cos\theta_z d\phi}
  &=&
  T n_t 
  \int_0^1 dy
  \left(
  \left.
  \frac{E_{\nu_\alpha}^{\rm {NC}}(E_{\rm sh}, y)}{E_{\rm sh}}
  V_{{\rm eff}, \nu_\alpha}^{\rm NC}(E_{\rm sh}, \cos\theta_z)
  \frac{d\sigma_{\nu_\alpha {\rm w}}^{\rm NC}(E_\nu, y)}{dy}
  \Phi^{\rm det}_{\nu_\alpha}(E_\nu,\cos\theta_z,\phi)
  \right\vert_{E_\nu = E_{\nu_\alpha}^{\rm NC}(E_{\rm sh}, y)}
  \right.
  \nonumber \\
  &&
  \qquad\qquad\quad
  \left.
  +~
  {\rm NC} \to {\rm CC}
  \right)
   \;,
 \end{eqnarray}
\end{widetext}
where $d\sigma_{\nu_\alpha {\rm w}}^{\rm NC}/dy$ is the cross section for interaction with water, made up of 10 protons and 8 neutrons, and $n_t$ is the number density of water molecules in ice.  The event rate due to $\bar{\nu}_\alpha$ is the same as \equ{spectrum_true}, but changing $\Phi^{\rm det}_{\nu_\alpha} \to \Phi^{\rm det}_{\bar{\nu}_\alpha}$, $d\sigma_{\nu_\alpha}^{\rm NC}/dy \to d\sigma_{\bar{\nu}_\alpha}^{\rm NC}/dy$, and $d\sigma_{\nu_\alpha}^{\rm NC}/dy \to d\sigma_{\bar{\nu}_\alpha}^{\rm CC}/dy$.  At these energies the cross sections for $\nu_\alpha$ and $\bar{\nu}_\alpha$ are nearly indistinguishable; see \Refe~\cite{Bertone:2016nfn} and Fig.~3 in \Refe~\cite{Valera:2022ylt}.  Equation~(\ref{equ:spectrum_true}) generalizes the original procedure in \Refe~\cite{Valera:2022ylt} by allowing the flux and the event rate to vary not only with zenith angle, $\theta_z$, but also with azimuth, $\phi$.  This allows our analysis to be sensitive to an excess of UHE neutrinos from the decay of DM towards the GC.

As in \Refe~\cite{Valera:2022ylt}, we smear the event rate using the detector energy and angular resolution, and use for our forecasts the event rate in terms of the reconstructed shower energy, $E_{\rm sh}^{\rm rec}$, and reconstructed direction, $\Omega^{\rm rec}$, \ie,
\begin{eqnarray}
 \frac{d^2 N_{\nu_\alpha}}{dE^\mathrm{rec}_\mathrm{sh}d\Omega^\mathrm{rec}}
 &=&
 \int dE_{\rm sh} 
 \int d\Omega 
 \frac{d^2 N_{\nu_\alpha}(E_{\rm sh}, \theta_z, \phi)}{dE_\mathrm{sh}d\Omega} 
 \nonumber \\
 &&
 \times~
 R_{E_\mathrm{sh}}(E^\mathrm{rec}_\mathrm{sh},E_\mathrm{sh}) R_{\Omega}(\mathbf{n}^\mathrm{rec},\mathbf{n}) \;,
\end{eqnarray}
where $d\Omega = \sin \theta_z d\theta_z d\phi$ and $d\Omega^{\rm rec} = \sin \theta_z^{\rm rec} d\theta_z^{\rm rec} d\phi^{\rm rec}$ are the real and reconstructed differential solid angles, and $\mathbf{n}$ and $\mathbf{n}^\mathrm{rec}$ are the real and reconstructed shower directions.  We model the energy resolution via a Gaussian function in $\epsilon \equiv \log_{10}(E_{\rm sh}^{\rm rec}/E_{\rm sh})$, \ie,
\begin{equation}
 R_{E_\mathrm{sh}}(E^\mathrm{rec}_\mathrm{sh},E_\mathrm{sh})
 =
 \sqrt{\frac{2}{\pi}}
 \frac{
 \exp\left[
 -\frac{(E^\mathrm{rec}_\mathrm{sh}-E_\mathrm{sh})^2}{2\sigma_{E_\mathrm{sh}}^2}
 \right]}{\sigma_{E_\mathrm{sh}}\left[1+\mathrm{Erf}\left(\frac{E^\mathrm{rec}_\mathrm{sh}}{\sqrt{2}\sigma_{E_\mathrm{sh}}}\right)\right]} \;,
\end{equation}
where $\sigma_{E_{\rm sh}} = 10^{\sigma_\epsilon} E_{\rm sh}$. 
As baseline, we fix $\sigma_\epsilon = 0.1$, based on simulations performed for UHE neutrino radio-detection at the RNO-G neutrino telescope~\cite{Aguilar:2021uzt}, which we take as representative of IceCube-Gen2, too.   We model the angular resolution via a Gaussian function of the angle between true and reconstructed direction, \ie,
\begin{equation}
 R_{\Omega}(\mathbf{n}^\mathrm{rec},\mathbf{n})
 =
 \frac{\sigma^2}{2\pi(1-e^{-2/\sigma^2})}
 \exp\left(
 \frac{\mathbf{n}\cdot\mathbf{n}_\mathrm{rec}}{\sigma^2}
 \right) \;,
\end{equation}
with a common width of $\sigma_{\theta_z} = \sigma_{\phi} =  \equiv \sigma_\Omega$ in zenith and azimuth.  As baseline, we fix $\sigma_\Omega = 3^\circ$, similar to what \Refe~\cite{Valera:2022ylt} adopted for the zenith-angle resolution.  References~\cite{Valera:2022ylt, Valera:2022wmu} explored the effect of varying the energy and angular resolution on the event rate.

To produce our forecasts, we use the all-flavor event rate of $\nu_\alpha$ and $\bar{\nu}_\alpha$, \ie,
\begin{equation}
 \label{equ:diff_event_rate_total}
 \frac{d^2 N_\nu} {dE_{\rm sh}^{\rm rec} d\Omega^{\rm rec}}
 =
 \sum_{\alpha=e,\mu,\tau}
 \left(
 \frac{d^2 N_{\nu_\alpha}} {dE_{\rm sh}^{\rm rec} d\Omega^{\rm rec}}
 +
 \frac{d^2 N_{\bar{\nu}_\alpha}} {dE_{\rm sh}^{\rm rec} d\Omega^{\rm rec}}
 \right) \;.
\end{equation}
Conservatively, we do not assume that radio-detection at IceCube-Gen2 will be able to distinguish between events initiated by different flavors; however, there is promising ongoing work in this direction~\cite{Garcia-Fernandez:2020dhb, Stjarnholm:2021xpj, Glaser:2021hfi}.


\subsection{Expected event rates}
\label{sec:detection_rates}

\begin{figure}
 \includegraphics[width=\columnwidth]{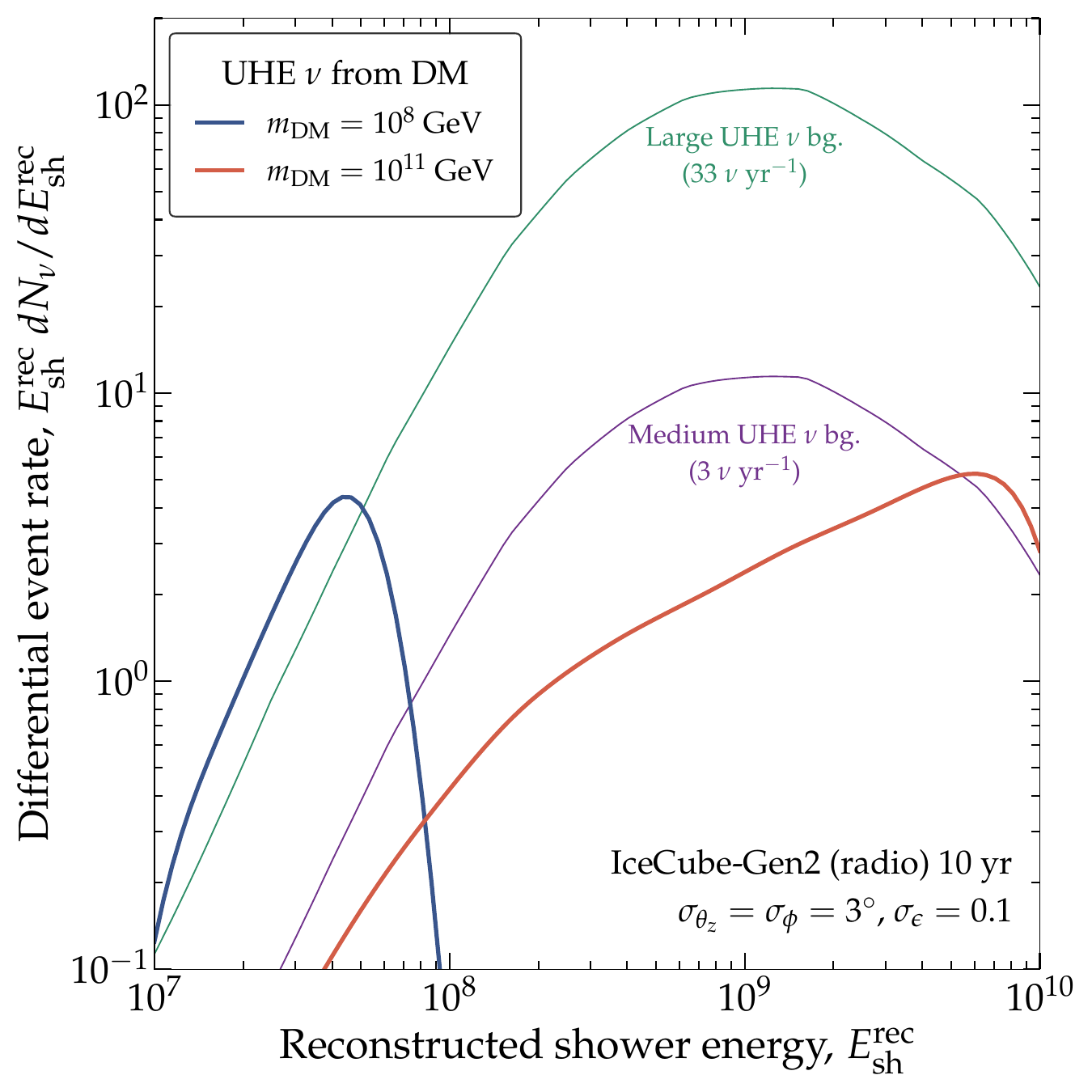}
 \caption{\textbf{\textit{Differential rate of neutrino-induced events in the radio array of IceCube-Gen2.}}  We show results obtained for the same illustrative choices of the flux from DM decay and the non-DM background flux as in \figu{diffuse_fluxes}.  The event rate is computed using \equ{diff_event_rate_total}, using baseline choices of the detector angular and (logarithmic) energy resolution, $\sigma_{\Omega} = 3^\circ$ and $\sigma_{\epsilon} = 0.1$.  This figure shows the direction-averaged energy distribution of events; \figu{sky_map_evrate} shows the angular event distribution.  See Sec.~\ref{sec:detection} for details.}\label{fig:diff_event_rate}
\end{figure}

\begin{figure}
 \includegraphics[width=\columnwidth]{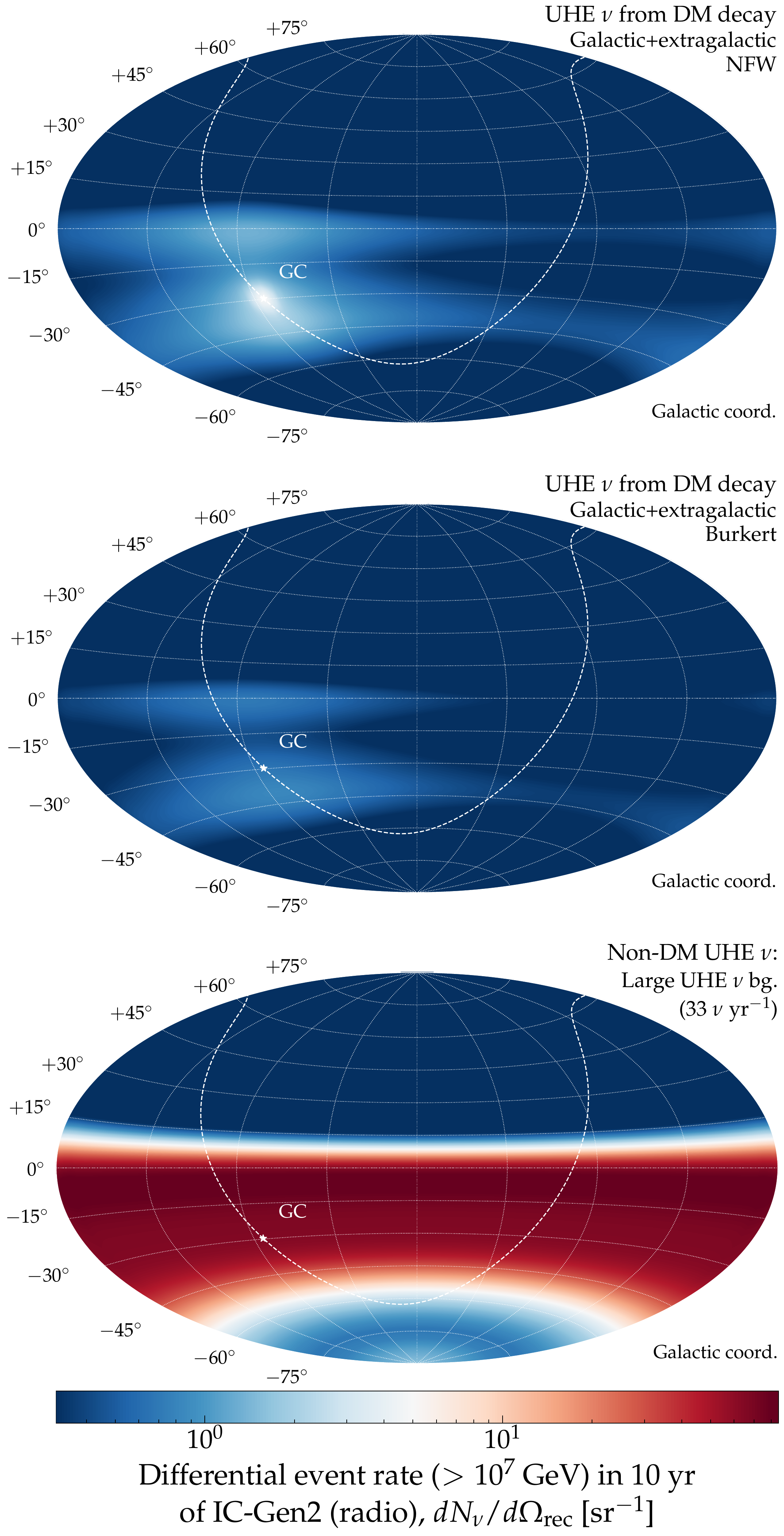}
 \caption{\textbf{\textit{Angular distribution of detected events in the radio array of IceCube-Gen2.}}  We use the same illustrative fluxes as in \figu{sky_map_flux}. {\it Top:} Neutrinos from DM decay for an NFW profile. {\it Center:} Same for a Burkert profile.  {\it Bottom:} Neutrinos from the non-DM isotropic background of UHE neutrinos. The angular dependence is due both to the angular distribution of the flux (\figu{sky_map_flux}) and the angular response of the detector (Fig.~13 in \Refe~\cite{Valera:2022ylt}).   The event rate is computed using \equ{diff_event_rate_total}, using baseline choices of the detector angular and (logarithmic) energy resolution, $\sigma_{\Omega} = 3^\circ$ and $\sigma_{\epsilon} = 0.1$.  This figure shows the energy-integrated angular distribution of events; \figu{diff_event_rate} shows the energy distribution.  See Sec.~\ref{sec:detection} for details.}\label{fig:sky_map_evrate}
\end{figure}

Using the methods above, we compute event rates for the flux of UHE neutrinos from DM decay and for the non-DM background flux of UHE neutrinos.  For the former, the event rate depends on the DM mass and lifetime.  For the latter, it depends on our choice of background flux.  As illustration, below we show event rates for the fluxes in \figu{diffuse_fluxes}; later, when producing results, we compute event rates for many more cases.

Figure~\ref{fig:diff_event_rate} shows the all-sky differential event rate in reconstructed shower energy.  The event energy spectra reflect the features of the underlying neutrino energy spectra in \figu{diffuse_fluxes}, though smoothed out by the detector energy resolution, which complicates distinguishing between them in our forecasts later (see also Sec.~\ref{sec:fluxes_astro_vs_dm}).  Further, while the neutrino energy spectrum from DM decay with $m_\mathrm{DM} = 10^8$~GeV and the spectrum of the large non-DM benchmark flux are comparable in \figu{diffuse_fluxes}, in \figu{diff_event_rate} the event rate for the former is appreciably smaller than that for the latter.  This is because the effective volume falls at low energies, where the spectrum from DM decay peaks; see Sec.~\ref{sec:detection_simulation}.

Figure~\ref{fig:sky_map_evrate} shows sky maps of the angular distribution of the energy-integrated event rate, for neutrinos from DM decay and for the background neutrinos.  The angular distribution of events is anisotropic, even when it is due to an isotropic neutrino flux, like the background flux.  The radio array of IceCube-Gen2 is mostly sensitive to zenith angles between $45^\circ$ and $90^\circ$.  At larger zenith angles, Earth attenuation strongly reduces the chances of neutrino detection, whereas at smaller zenith angles the effective volume is smaller.  For this reason, most of the events come from declinations between $-45^\circ$ and $0^\circ$.  (The two bright zenith bands in the skymaps, easily visible for neutrinos from DM decay, are due to features in the response of the two types of antennas that the radio array is made of; see Fig.~12 in \Refe~\cite{Valera:2022ylt}.)

The sky maps in \figu{sky_map_evrate} illustrate the combined effect of the three sources of angular dependence in our calculation: from the neutrino flux itself, from the propagation of neutrinos through the Earth, and from the detector effective volume.  The latter two, together with the angular resolution of the detector, smooth out any natural anisotropy in the neutrino flux.  Nevertheless, \figu{sky_map_evrate} shows that the excess of neutrinos from DM decay towards the GC survives into the angular distribution of events, though it is more spread out.  The excess is more concentrated for the NFW profile than for the Burkert profile, reflecting their fluxes from \figu{sky_map_evrate}.   


\section{Discovery prospects for dark matter decay}
\label{sec:discovery}

\textit{The decay of heavy DM into UHE neutrinos may be discovered even in the presence of sizable non-DM neutrino backgrounds, by using the angular distribution of detected events, in 10~years of exposure of the radio array of IceCube-Gen2 (\figu{discovery_prospects}).  However, a puffy Galactic DM density profile may weaken the discovery prospects (\figu{discovery_prospects_nfw_vs_burkert}).  Upon discovery, the DM mass and lifetime, and the flux of neutrinos from its decay, may be accurately and precisely measured by using also the energy distribution of events (Figs.~\ref{fig:reconstructed} and \ref{fig:flux_reconstructed}).}


\subsection{Overview}
\label{sec:discovery_overview}

The distinct angular distribution of UHE neutrinos from DM decay---peaked towards the GC---provides a smoking-gun signature of their origin when compared to the isotropic flux of astrophysical and cosmogenic neutrinos.  Yet, so far, the usefulness of this difference has gone underused or ignored in forecasts of searches for DM decay in UHE neutrino telescopes; see, \eg, \Refe~\cite{Chianese:2021jke}.  In contrast, our methods embrace it.  Unlike previous forecasts, we use this angular difference to not only claim the discovery of UHE neutrinos with a {\it possible} origin in DM decay, but to {\it assert} their DM origin in the presence of a non-DM neutrino background, \ie, to firmly discover UHE neutrinos from DM decay.  

However, a sensible DM discovery claim requires a sufficiently large excess towards the GC. Added to that, if there is a large isotropic background of non-DM neutrinos, it could wash out the excess of neutrinos from DM decay, weakening the discovery claim.

In our forecasts below, we quantify the above statements in two ways. First (Sec.~\ref{sec:discovery_prospects}), we find the values of the mass and lifetime of DM needed to discover UHE neutrinos from its decay, in the presence of a non-DM neutrino background.  For this, we use only the angular distribution of detected events in the radio array of IceCube-Gen2.  Second (Sec.~\ref{sec:discovery_measuring}), in the event of discovery, we illustrate the accuracy with which the DM mass and lifetime could be measured.  For this, we use the joint angular and energy distribution of events.


\subsection{Discovery prospects}
\label{sec:discovery_prospects}

\begin{figure}[t]
 \includegraphics[width=\columnwidth]{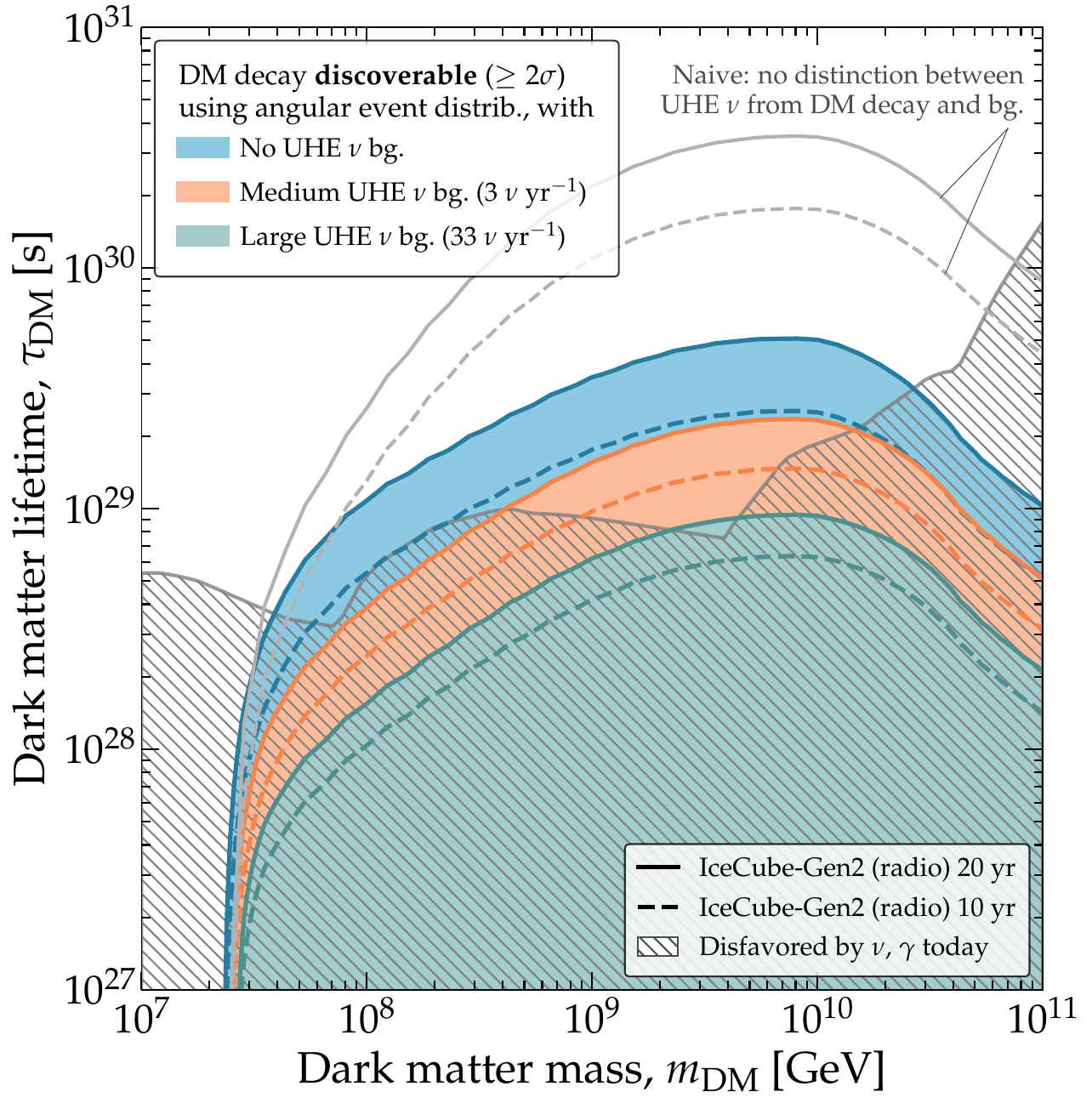}
 \caption{\textit{\textbf{Discovery prospects of UHE neutrinos from DM decay in the radio array of IceCube-Gen2.}}  Forecasts are for three benchmark choices of the background flux of non-DM UHE neutrinos (\ie, cosmogenic): null, medium, and large (Sec.~\ref{sec:fluxes_astro}), based on the cosmogenic neutrino flux by Bergman \& van Vliet~\cite{Anker:2020lre} (\figu{diffuse_fluxes}).  The forecasts use only the angular distribution of events, summed over all reconstructed energies $E_{\rm sh}^{\rm rec} \geq 10^7$~GeV.  They use baseline choices of the detector angular and (logarithmic) energy resolution, $\sigma_{\Omega} = 3^\circ$ and $\sigma_{\epsilon} = 0.1$; see Sec.~\ref{sec:detection_simulation}.  This figure assumes the NFW density profile for Galactic DM.  Existing lower limits on the DM lifetime are the same as in \figu{summary}.
 In this figure, results are for discovery at $\geq 2 \sigma$; see \figu{discovery_prospects_3sigma} for results at $\geq 3\sigma$.  \textit{Even in the presence of a medium-sized background, DM discovery might be possible.} See Sec.~\ref{sec:discovery_prospects} for details.}\label{fig:discovery_prospects}
\end{figure}

We forecast the regions of DM mass and lifetime where UHE neutrinos from DM decay could be discovered.  Figure~\ref{fig:discovery_prospects} (also \figu{discovery_prospects_3sigma}) shows our results.


\subsubsection{Statistical methods}
\label{sec:discovery_prospects_stat}

We produce discovery forecasts by analyzing projected samples of detected events.  To be conservative, we use only their angular distribution, summed over all energies, since it is in it that the critical difference between the flux of neutrinos from DM decay and the non-DM neutrino background manifests (Sec.~\ref{sec:fluxes_astro_vs_dm}).  Later (Sec.~\ref{sec:bounds}), we derive upper limits using also their energy distribution.  We build our forecasts using the maximum likelihood technique and report mean discovery prospects based on Asimov data samples.

Each event sample is the sum of events due to neutrinos from DM decay, dependent on the DM mass, $m_\mathrm{DM}$, and lifetime, $\tau_\mathrm{DM}$, and events from the background of non-DM neutrinos, rescaled by a flux normalization, $\mathcal{N}_\Phi$, \ie,
\begin{equation}
 \label{equ:generation_event_rate}
 \frac{dN_\nu(\boldsymbol{\vartheta})}{d\Omega^\mathrm{rec}}
 =
 \frac{dN^\mathrm{DM}_\nu\left(m_\mathrm{DM}, \tau_\mathrm{DM}\right)}{d\Omega^\mathrm{rec}}
 +
 \mathcal{N}_\Phi\frac{dN^\mathrm{bg}_\nu}{d\Omega^\mathrm{rec}} \;,
\end{equation}
where $\boldsymbol{\vartheta} \equiv \left\{m_\mathrm{DM}, \tau_\mathrm{DM}, \mathcal{N}_\Phi\right\}$.  For the non-DM background, we show forecasts obtained under the three benchmark scenarios presented in Sec.~\ref{sec:fluxes_astro}: a large flux set to the cosmogenic flux by Bergman \& van Vliet~\cite{Anker:2020lre}, a medium flux that is 10\% of that, and a null background; see \figu{diffuse_fluxes}.  The large and medium benchmark backgrounds have the same angular distribution of events (see \figu{sky_map_evrate} and also Fig.~4 in \Refe~\cite{Valera:2022wmu}); they only differ in the total number of events (\figu{diff_event_rate}).

Based on a projected event sample, we compare two hypotheses: the DM hypothesis, where the angular distribution of events is best explained by the presence of a DM decay component on top of a background non-DM component {\it vs.}~the null hypothesis, where it is best explained by the presence of only the background non-DM component. In both cases, the background flux normalization, $\mathcal{N}_\Phi$, is left free to vary to best fit the data.  Hence, when analyzing a simulated event sample, we obtain results that do not require prior knowledge of the true background flux.  We do this separately for each of the three above choices of the simulated background flux, and for a wide range of true values of the DM and lifetime.  

First, for a particular choice of the true DM mass and lifetime, $m_{\rm DM}$ and $\tau_{\rm DM}$, and using the true value of the flux normalization, $\mathcal{N}_{\Phi} = 1$, we compute the projected {\it observed} sample of $N_{\rm evts}$ events, each with reconstructed direction $\Omega_i^{\rm rec}$ sampled from the distribution $dN_\nu(m_{\rm DM}, \tau_{\rm DM}, \mathcal{N}_\Phi = 1) / d\Omega^{\rm rec}$.  Later, we use a test statistic that is averaged over all possible random realizations of the number of events and of the distribution of reconstructed directions of the events.

Then, based on this observed sample, we evaluate an unbinned likelihood function at different test values of the model parameters, $\boldsymbol{\vartheta}^\prime \equiv (m_{\rm DM}^\prime, \tau_{\rm DM}^\prime, \mathcal{N}_\Phi^\prime)$, \ie,
\begin{equation}
 \label{equ:likelihood_dm}
 \mathcal{L}
 (\boldsymbol{\vartheta}^\prime; 
 \left\{\Omega_{i}^{\rm rec}\right\})
 =
 e^{-N_\nu(\boldsymbol{\vartheta^\prime)}} 
 \prod_{i=1}^{N_{\rm evts}}
 \left.
 \frac{dN_\nu(\boldsymbol{\vartheta}^\prime)}{d\Omega^\mathrm{rec}}
 \right|_{\Omega^\mathrm{rec}_i} \;,
\end{equation}
where $N_\nu(\boldsymbol{\vartheta}^\prime) \equiv \int (dN_\nu(\boldsymbol{\vartheta}^\prime) / d\Omega^\mathrm{rec}) d\Omega^\mathrm{rec}$ is the all-sky event rate.  In the comparison, we let the test values of $m_{\rm DM}^\prime$, $\tau_{\rm DM}^\prime$, and $\mathcal{N}_\Phi^\prime$ float as free parameters, as they would in a test based on real experimental data.  Under the null hypothesis, where there is no DM decay contribution because DM is stable (\ie, $\tau_{\rm DM} \to \infty$), the likelihood reduces to
\begin{equation}
 \label{equ:likelihood_bg}
 \mathcal{L}_{\rm bg}(\mathcal{N}_\Phi^\prime;
 \left\{ \Omega_i^{\rm rec} \right\})
 \equiv
 \lim_{\tau_{\rm DM}^\prime \to \infty}
 \mathcal{L}(\boldsymbol{\vartheta}^\prime;
 \left\{ \Omega_i^{\rm rec} \right\}) \;,
\end{equation}
where the right-hand side no longer depends on the test values $m_{\rm DM}^\prime$ and $\tau_{\rm DM}^\prime$.

For a specific choice of the true values of $m_{\rm DM}$ and $\tau_{\rm DM}$, the test statistic depends on the angular distribution of the associated random observed event sample.  To account for the possible different realizations of the observed sample, we average the logarithm of the likelihood functions, Eqs.~(\ref{equ:likelihood_dm}) and (\ref{equ:likelihood_bg}), over all possible realizations.  The probability to observe a total number of $N_{\rm evts}$, over the full sky, $\mathcal{P}(N_{\rm evts} \vert N_\nu)$, is given by a Poisson distribution with a mean value equal to the mean all-sky event rate, $N_\nu \equiv N_\nu(m_{\rm DM}, \tau_{\rm DM}, \mathcal{N}_\Phi = 1)$.  The probability to sample an event with reconstructed direction $\Omega_i^{\rm rec}$ from this distribution is $\mathcal{P}(\Omega_i^{\rm rec}) \equiv (1/N_\nu) (dN_\nu/d\Omega^{\rm rec})\vert_{\Omega_i^{\rm rec}}$.  Thus, the likelihood function, \equ{likelihood_dm}, averaged over all possible realizations of the event sample, is
\begin{widetext}
\begin{eqnarray}
 \label{equ:likelihood_dm_avg}
 \langle
 \ln \mathcal{L}
 (\boldsymbol{\vartheta}^\prime)
 \rangle_{m_{\rm DM}, \tau_{\rm DM}}
 &=&
 \sum_{N_{\rm evts}=0}^\infty
 \mathcal{P}
 \left[
 N_{\rm evts} \vert N_\nu(m_{\rm DM}, \tau_{\rm DM}, \mathcal{N}_\Phi=1)
 \right]
 \\
 && \nonumber
 \times 
 \int d\Omega_1^{\rm rec}
 \cdots
 \int d\Omega_{N_{\rm evts}}^{\rm rec}
 \prod_{i=1}^{N_{\rm evts}}
 \mathcal{P}(\Omega_i^{\rm rec} \vert m_{\rm DM}, \tau_{\rm DM}, \mathcal{N}_\Phi=1)
 \ln \mathcal{L}
 \left[
 \boldsymbol{\vartheta}^\prime, 
 \left\{\Omega_j^{\rm rec}\right\}_{j=1}^{N_{\rm evts}}
 \right] \;,
\end{eqnarray}
\end{widetext}
and, similarly, the average log-likelihood under the null hypothesis, in which only background non-DM neutrinos are present, \equ{likelihood_bg}, is
\begin{equation}
 \label{equ:likelihood_avg_bg}
 \langle
 \ln \mathcal{L}_{\rm bg}(\mathcal{N}_\Phi^\prime)
 \rangle_{m_{\rm DM}, \tau_{\rm DM}}
 \equiv
 \lim_{\tau_{\rm DM}^\prime \to \infty} 
 \langle
 \ln \mathcal{L}(\boldsymbol{\vartheta}^\prime) 
 \rangle_{m_{\rm DM}, \tau_{\rm DM}} \;,
\end{equation}
We average the log-likelihood, rather than the likelihood, to prevent the averaging procedure from prescribing exceedingly large averaging weights to  random realizations that have associated large likelihood values. This corresponds to obtaining the results for an Asimov data sample~\cite{Cowan:2010js}, in which the observed distribution of events exactly coincide with the expected one.

To compare the two hypotheses, we use as a test statistic the average log-likelihood ratio, \ie,
\begin{eqnarray}
 \label{equ:test_statistic}
 \langle
 \Lambda(m_{\rm DM}, \tau_{\rm DM})
 \rangle
 &=&
 \mathrm{min}_{\mathcal{N}_\Phi^\prime}
 \left[
 -2
 \langle
 \ln\mathcal{L}_\mathrm{bg}(\mathcal{N}_\Phi^\prime)
 \rangle_{ m_{\rm DM}, \tau_{\rm DM}}
 \right]
 \nonumber \\
 && -~
 \mathrm{min}_{\boldsymbol{\vartheta}^\prime}
 \left[
 -2
 \langle
 \ln\mathcal{L}(\boldsymbol{\vartheta}^\prime)
 \rangle_{ m_{\rm DM}, \tau_{\rm DM}}
 \right] \;.
\end{eqnarray}
According to Wilks' theorem~\cite{Wilks:1938dza}, in the asymptotic limit of a large data sample, this quantity follows a $\chi^2$ distribution with two degrees of freedom, corresponding to the difference between the dimensions of the parameter spaces of the two competing hypotheses.  We adopt it in our forecasts since, for 10 and 20 years of detector exposure, and for the neutrino fluxes that we adopt, they are based on a large number of events.  Hence, below, when $\langle \Lambda \rangle > 6$, we claim discovery of DM neutrinos at the $2\sigma$ confidence level (C.L.); when $\langle \Lambda \rangle > 11.5$, we claim it at $3\sigma$~C.L.

\begin{figure}[t]
 \includegraphics[width=\columnwidth]{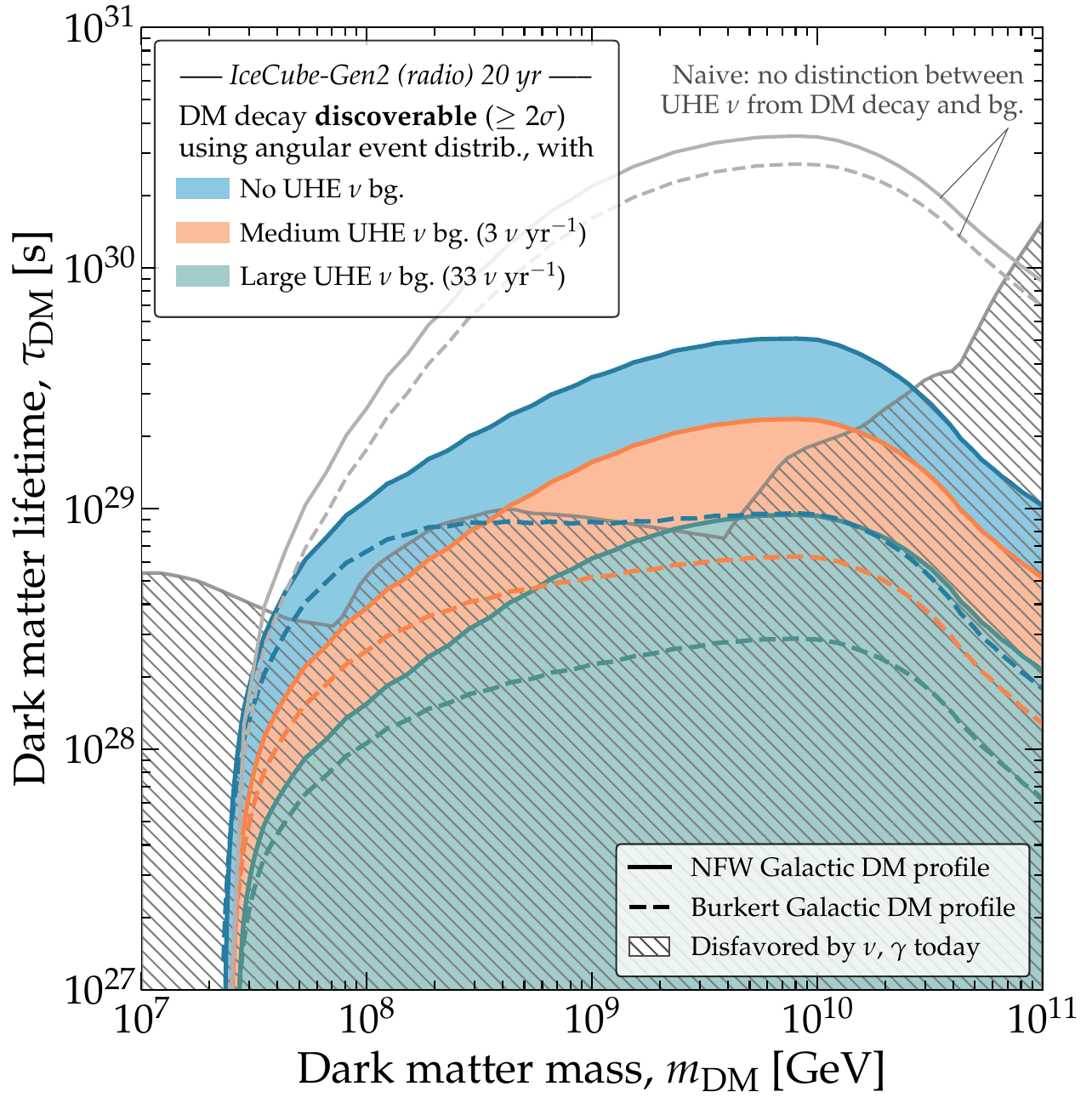}
 \caption{\textit{\textbf{Discovery prospects of UHE neutrinos from DM decay for NFW vs.~Burkert Galactic DM profiles.}}  Same as \figu{discovery_prospects}, but comparing results obtained using the NFW {\it vs.}~Burkert Galactic DM density profiles (\figu{dm_profiles}).  Results are for discovery at $\geq 2 \sigma$ after 20 years; see \figu{discovery_prospects_burkert} for more results using the Burkert profile.  Existing lower limits on the DM lifetime are the same as in \figu{summary}, and computed assuming the NFW profile~\cite{Arguelles:2022nbl}.  \textit{If the Galactic DM profile is ``puffy'', like the Burkert profile, discovery of DM becomes barely feasible, and only if there is no isotropic non-DM UHE neutrino background.} See Sec.~\ref{sec:discovery_prospects} for details.}\label{fig:discovery_prospects_nfw_vs_burkert}
\end{figure}


\subsubsection{Results}

Figure~\ref{fig:discovery_prospects} shows the regions of DM mass and lifetime, obtained with the above methods, where DM can be discovered at $\geq 2\sigma$.  We show results for our three benchmark choices of non-DM UHE neutrino background flux (Sec.~\ref{sec:fluxes_astro})---null, medium, and large.  The results in \figu{discovery_prospects} convey three key messages.

First, while a large part of the parameter space in \figu{discovery_prospects} is already disfavored by present-day neutrino and gamma-ray searches, upcoming UHE neutrino telescopes will extend the search to longer DM lifetimes, roughly above $10^{29}$~s, that are unreachable with present-day experiments.  This aspect of our results agrees with previous works; see, \eg, \Refes~\cite{Chianese:2021jke, Arguelles:2022nbl}).  Yet, for the first time, we fortify the claim by showing that it holds even in the presence of a medium-sized isotropic background flux of non-DM UHE neutrinos.

Second, \figu{discovery_prospects} shows the significant difference between \textit{detecting} events, which a priori may or may not be due to DM decay, and \textit{claiming} that those events are produced by DM decay. Earlier forecasts of UHE neutrino detection from DM decay~\cite{Chianese:2021jke} had only investigated the maximum lifetime needed to detect neutrinos from DM decay, without identifying their origin. We show a version of these forecasts in \figu{discovery_prospects}, generated by demanding that DM decay yields at least one event over the full sky, background-free, with a probability larger than 95\% (or 99.7\% in \figu{discovery_prospects_3sigma}). As expected, this weaker criterion leads to overly long lifetimes being discoverable: compared to our results using the angular distribution of events, lifetimes longer by at least one order of magnitude could be detected without actually leading to a DM discovery. Thus, hereafter the main observations and conclusions of our work are based exclusively on forecasts made using the angular---and energy (in Secs.~\ref{sec:discovery_measuring} and \ref{sec:bounds})---distribution of events.

Third, \figu{discovery_prospects} shows that, while the discovery prospects are best when background-free, as expected, the presence of a medium-size background only degrades the reach of discoverable DM lifetimes by a factor of about 2.  In other words, \textbf{\textit{UHE neutrinos retain the potential to reveal DM decay even in the presence of a sizable isotropic background of non-DM origin}}.  This is true also for discovery at $3\sigma$; see \figu{discovery_prospects_3sigma}.

Figure~\ref{fig:discovery_prospects_nfw_vs_burkert} shows that, however, our prospects for discovery of DM decay are contingent on the Galactic DM density profile being cuspy, \ie, markedly pronounced towards the GC.  Swapping the cuspy NFW profile for the puffy Burkert profile reduces the region amenable for discovery by about one order of magnitude, pushing it into the region of DM mass and lifetime that is already disfavored, and rendering discovery all but unfeasible.  A subtle point is that the present-day disfavored region shown in \figu{discovery_prospects_nfw_vs_burkert} was computed, in \Refe~\cite{Arguelles:2022nbl}, assuming the NFW profile. Assuming the Burkert instead would push down the disfavored region, leaving slightly more room for discovery under the Burkert profile; we do not attempt this recalculation here.  Yet, the bottom line holds: for realistic choices of the detector angular resolution, like the $\sigma_\Omega = 3^\circ$ that we adopt, \textbf{\textit{the discovery of DM decay into UHE neutrinos will be likely only if the Galactic DM density profile is cuspy}}.


\subsection{Measuring dark matter mass and lifetime}
\label{sec:discovery_measuring}

In the event of the discovery of DM decay into UHE neutrinos, we forecast how well the DM mass and lifetime could be inferred.  Figure~\ref{fig:reconstructed} shows our results.


\subsubsection{Statistical methods}
\label{sec:discovery_measuring_stat}

Above (Sec.~\ref{sec:discovery_prospects}), we showed that to claim the discovery of DM decay into UHE neutrinos it was enough to use the angular distribution of events.  In the event of discovery, inferring the DM lifetime also relies mainly on the angular distribution events; concretely, as before, on its excess towards the GC.  Because the DM lifetime determines the normalization of the neutrino flux from DM decay, its value can be inferred from the magnitude of the excess.  However, inferring the DM mass requires the energy distribution of events, too.  Using it allows us to infer the DM mass by looking for the distinct bump-like feature imprinted by DM decay on the neutrino spectrum, which peaks at $E_\nu = m_{\rm DM}/2$; see Figs.~\ref{fig:diffuse_fluxes} and \ref{fig:diff_event_rate}.  Thus, below, we extend the statistical methods from Sec.~\ref{sec:discovery_prospects_stat} to include also the energy of the detected events.

First, we generate the {\it true} event sample, \ie, the one that we assume will be detected.  We use a procedure similar to the one we used to compute discovery prospects (Sec.~\ref{sec:discovery_prospects}), but extended to included also the energy distribution of events. To illustrate our method, we choose $m_{\rm DM}=3.5\times 10^9$~GeV and $\tau_{\rm DM}=1.19\times 10^{29}$~s as true values; these are representative of the discoverable region under our benchmark medium non-DM in \figu{discovery_prospects}.  We use for the non-DM background UHE neutrino flux the same three benchmarks (Sec.~\ref{sec:fluxes_astro}) that we used earlier to make discovery forecasts---null, medium, and large.  In analogy to \equ{generation_event_rate}, the true distribution is
\begin{equation}
 \label{equ:generation_event_rate_recon}
 \frac{dN_\nu^{\rm true}(m_{\rm DM}, \tau_{\rm DM})}{dE_{\rm sh}^{\rm rec} d\Omega^\mathrm{rec}}
 =
 \frac{dN^\mathrm{DM}_\nu\left(m_\mathrm{DM}, \tau_\mathrm{DM}\right)}{dE_{\rm sh}^{\rm rec} d\Omega^\mathrm{rec}}
 +
 \frac{dN^\mathrm{bg}_\nu}{dE_{\rm sh}^{\rm rec} d\Omega^\mathrm{rec}} \;.
\end{equation}
From it, we randomly sample $N_{\rm evts}$ events, each with reconstructed energy $E_{{\rm sh}, i}^{\rm rec}$ and direction $\Omega_i^{\rm rec}$.  Later, for each choice of $m_{\rm DM}$ and $\tau_{\rm DM}$, we average our test statistic over all possible random realizations of the event samples.

Then we compare the true event sample to {\it test} event samples, generated for many different test values of DM mass and lifetime, in order to find which ones fit best.  To produce test event samples, we generalize what we did to compute discovery prospects and adopt a generic model of the non-DM background neutrino energy spectrum.  We parametrize it as a piecewise (pw) spectrum $\propto E_\nu^{-2}$, with three independent normalization constants, $\mathcal{N}_{\Phi, 1}$, $\mathcal{N}_{\Phi, 2}$, $\mathcal{N}_{\Phi, 3}$, in three decades of neutrino energy, from $10^7$~GeV to $10^{10}$~GeV.  For $\nu_\alpha$, this is
\begin{equation}
 \label{equ:flux_bg_pw}
 \Phi_{\nu_\alpha}^{\rm bg-pw}(E_\nu)
 =
 \frac{f_{\alpha, \oplus}}{2 E_\nu^2} \times
 \left\{
  \begin{array}{ll}
   \mathcal{N}_{\Phi, 1}, & 10^7 \leq E_\nu/{\rm GeV} < 10^8 \\
   \mathcal{N}_{\Phi, 2}, & 10^8 \leq E_\nu/{\rm GeV} < 10^9 \\
   \mathcal{N}_{\Phi, 3}, & 10^9 \leq E_\nu/{\rm GeV} \leq 10^{10} \\
  \end{array}
 \right. \;,
\end{equation}
where, for the flavor composition at Earth, we adopt the one from the canonical expectation of neutrino production via the  pion decay chain (Sec.~\ref{sec:fluxes_astro}), computed using recent best-fit values of the neutrino mixing parameters~\cite{Esteban:2020cvm, NuFit5.0}, $f_{e,\oplus} = 0.298$, $f_{\mu,\oplus} = 0.359$, and $f_{\tau,\oplus} = 0.342$, from \Refe~\cite{Song:2020nfh}, \ie, close to flavor equipartition (see also, \eg, \Refe~\cite{Bustamante:2015waa}).  Since we make forecasts for 10--20 years, when the values of the mixing parameters will likely be known precisely~\cite{Valera:2022ylt}, we neglect the small uncertainties in these predictions; see Eqs.~(7)--(9) in \Refe~\cite{Valera:2022ylt} . The fluxes of $\nu_\alpha$ and $\bar{\nu}_\alpha$ are identical; this is ensured by the factor of 2 in the denominator of \equ{flux_bg_pw}. 

We adopt this background flux model to analyze projected event samples using a phenomenological prescription of the non-DM neutrino background that is as agnostic as possible regarding the shape of its energy spectrum and its origin.  Our strategy resembles one that would be used by future analyses based on real experimental observations.  Indeed, similar flux models are used by the IceCube Collaboration to analyze present-day data~\cite{IceCube:2020wum, IceCube:2021uhz}. Using the piecewise background flux, we compute the associated differential event spectrum, $dN_\nu^{\rm bg-pw}/dE_{\rm sh}^{\rm rec}d\Omega^{\rm rec}$, using the methods from Sec.~\ref{sec:detection_simulation}.  Thus, the total differential test event spectrum is
\begin{eqnarray}
 \label{equ:test_event_rate_measure}
 \frac{dN_\nu^{\rm test}(\boldsymbol{\vartheta}^\prime)}{dE^\mathrm{rec}_\mathrm{sh}d\Omega^\mathrm{rec}}
 &=&
 \frac{dN^\mathrm{DM}_\nu(m_\mathrm{DM}^\prime,\tau_\mathrm{DM}^\prime)}{dE^\mathrm{rec}_\mathrm{sh}d\Omega^\mathrm{rec}}
 \\ \nonumber 
 && +
 \frac{dN_\nu^{\mathrm{bg-pw}}(\mathcal{N}_{\Phi, 1}^\prime, \mathcal{N}_{\Phi, 2}^\prime, \mathcal{N}_{\Phi, 3}^\prime)}{dE^\mathrm{rec}_\mathrm{sh} d\Omega^\mathrm{rec}} 
 \; ,
\end{eqnarray}
where now $\boldsymbol{\vartheta}^\prime \equiv (m_{\rm DM}^\prime, \tau_{\rm DM}^\prime, \mathcal{N}_{\Phi, 1}^\prime, \mathcal{N}_{\Phi, 2}^\prime, \mathcal{N}_{\Phi, 3}^\prime)$.

In analogy to \equ{likelihood_dm}, we evaluate an unbinned likelihood function at different test values of the model parameters, $\boldsymbol{\vartheta}^\prime$, \ie,
\begin{eqnarray}
 \label{equ:likelihood_mea}
 \mathcal{L}_{\rm mea}(\boldsymbol{\vartheta}^\prime, 
 \left\{ E_{{\rm sh}, i}^{\rm rec}, \Omega_i^{\rm rec} \right\})
 &=&
 e^{-N_\nu^{\rm test}(\boldsymbol{\vartheta}^\prime)}
 \\
 && \nonumber
 \times
 \prod_{i=1}^{N_\mathrm{evts}}
 \left.
 \frac{dN_\nu^{\rm test}(\boldsymbol{\vartheta}^\prime)}{dE^\mathrm{rec}_\mathrm{sh}d\Omega^\mathrm{rec}}
 \right\vert_{E_{{\rm sh}, i}^{\rm rec}, \Omega_i^{\rm rec}}
 \;,
\end{eqnarray}
where $N_\nu^{\rm test}$ is the all-sky number of events with energies $E_{\rm sh}^{\rm rec} \geq 10^7$~GeV.  As before (Sec.~\ref{sec:discovery_prospects}), for a specific choice of the true values of $m_\mathrm{DM}$ and $\tau_\mathrm{DM}$ and of the non-DM background neutrino flux, we average the logarithm of the above likelihood over all possibly random realizations of the observed event sample.  In analogy to \equ{likelihood_dm_avg}, this yields $\langle \ln \mathcal{L}_{\rm mea}(\boldsymbol{\vartheta}^\prime) \rangle_{m_{\rm DM}, \tau_{\rm DM}}$.  Finally, to infer the values of the DM mass and lifetime, we profile the likelihood over the test parameters and, in analogy to \equ{test_statistic}, define the test statistic
\begin{widetext}
\begin{eqnarray}
 \label{equ:test_statistic_recon}
 \langle \Lambda_{\rm mea}(m_\mathrm{DM},\tau_\mathrm{DM})\rangle
 &=&
 \mathrm{min}_{\mathcal{N}_{\Phi, 1}^\prime, \mathcal{N}_{\Phi, 2}^\prime, \mathcal{N}_{\Phi, 3}^\prime}
 \left[
 -2 \langle
 \ln \mathcal{L}_{\rm mea}
 (
 m_{\rm DM}^\prime = m_{\rm DM}, 
 \tau_{\rm DM}^\prime = \tau_{\rm DM}, 
 \mathcal{N}_{\Phi, 1}^\prime, 
 \mathcal{N}_{\Phi, 2}^\prime, 
 \mathcal{N}_{\Phi, 3}^\prime
 )
 \rangle_{m_{\rm DM}, \tau_{\rm DM}}
 \right]
 \nonumber \\
 &&
 - ~
 \mathrm{min}_{\boldsymbol{\vartheta}^\prime}
 \left[
 -2 \langle
 \ln \mathcal{L}_{\rm mea}
 (\boldsymbol{\vartheta}^\prime)
 \rangle_{m_{\rm DM}, \tau_{\rm DM}}
 \right] \;.
\end{eqnarray}
\end{widetext}
As in Sec.~\ref{sec:discovery_prospects}, based on Wilks' theorem, this test statistic follows a $\chi^2$ distribution with two degrees of freedom.  Below, we use it to infer allowed regions of $m_\mathrm{DM}$ and $\tau_\mathrm{DM}$ at difference confidence levels.


\subsubsection{Results}

\begin{figure}[t]
 \includegraphics[width=\columnwidth]{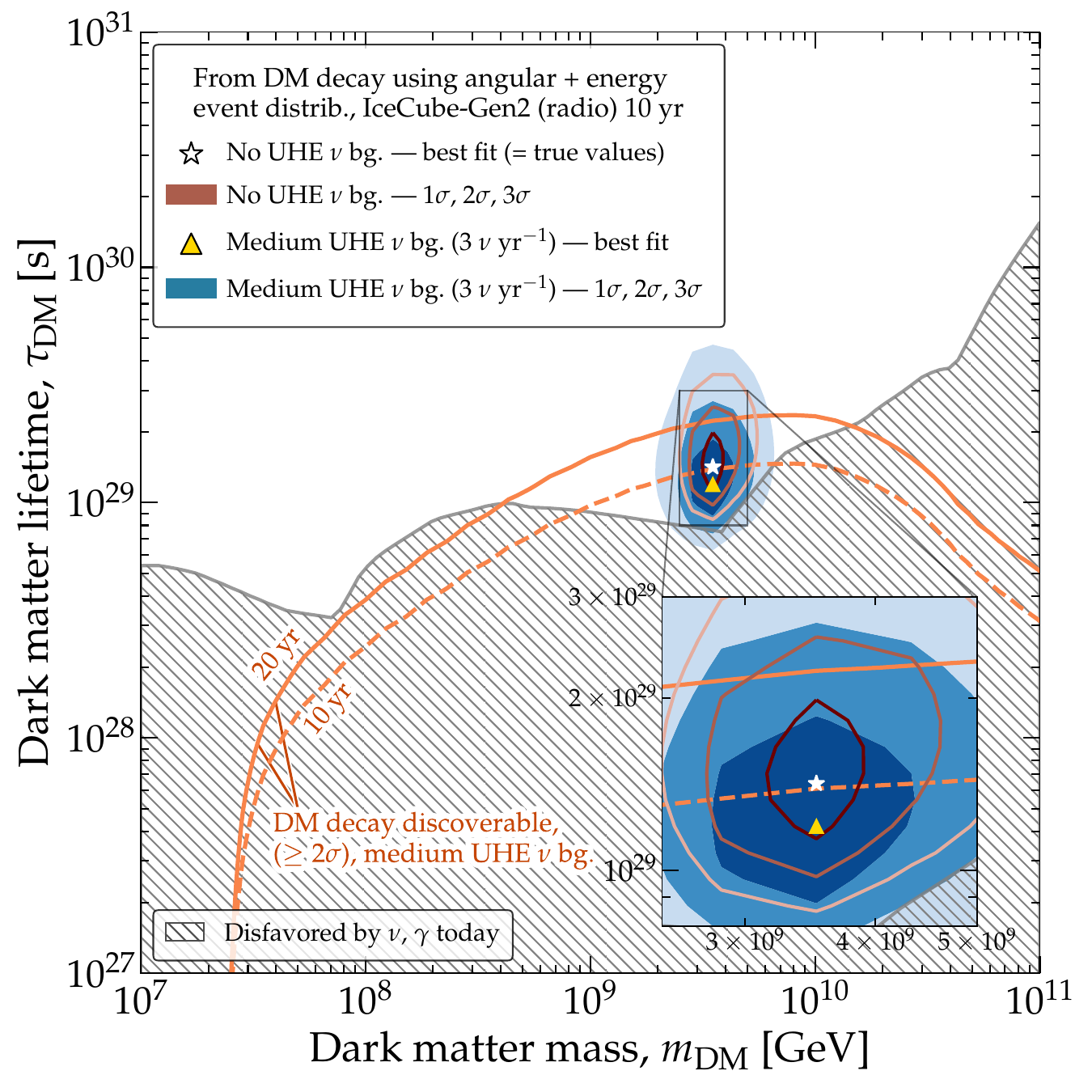}
 \caption{\textit{\textbf{DM mass and lifetime inferred from detecting UHE neutrinos from DM decay in the radio array of IceCube-Gen2.}}  Results are generated for an illustrative choice of the true values of the DM mass and lifetime, $m_{\rm DM} = 3.5 \times 10^9$~GeV and $\tau_{\rm DM} = 1.19 \times 10^{29}$~s, which are representative of what is discoverable with 10~years of detector exposure and allowed by present-day limits (\figu{discovery_prospects}).  Forecasts are for two choices of the background isotropic flux of non-DM UHE neutrinos: null and medium (Sec.~\ref{sec:fluxes_astro}), \ie, 10\% of the cosmogenic flux by Bergman \& van Vliet~\cite{Anker:2020lre} (\figu{diffuse_fluxes}).  The forecasts use the joint angular and energy distribution of events, and baseline choices of the detector angular and (logarithmic) energy resolution, $\sigma_{\Omega} = 3^\circ$ and $\sigma_{\epsilon} = 0.1$; see Sec.~\ref{sec:detection_simulation}.  This figure assumes the NFW density profile for Galactic DM. Existing lower limits on the DM lifetime are the same as in \figu{summary}.  \textit{In the event of DM discovery, the DM mass and lifetime may be measured with reasonable accuracy and precision.} See Sec.~\ref{sec:discovery_measuring} for details.}\label{fig:reconstructed}
\end{figure}

\begin{figure}[t]
 \includegraphics[width=\columnwidth]{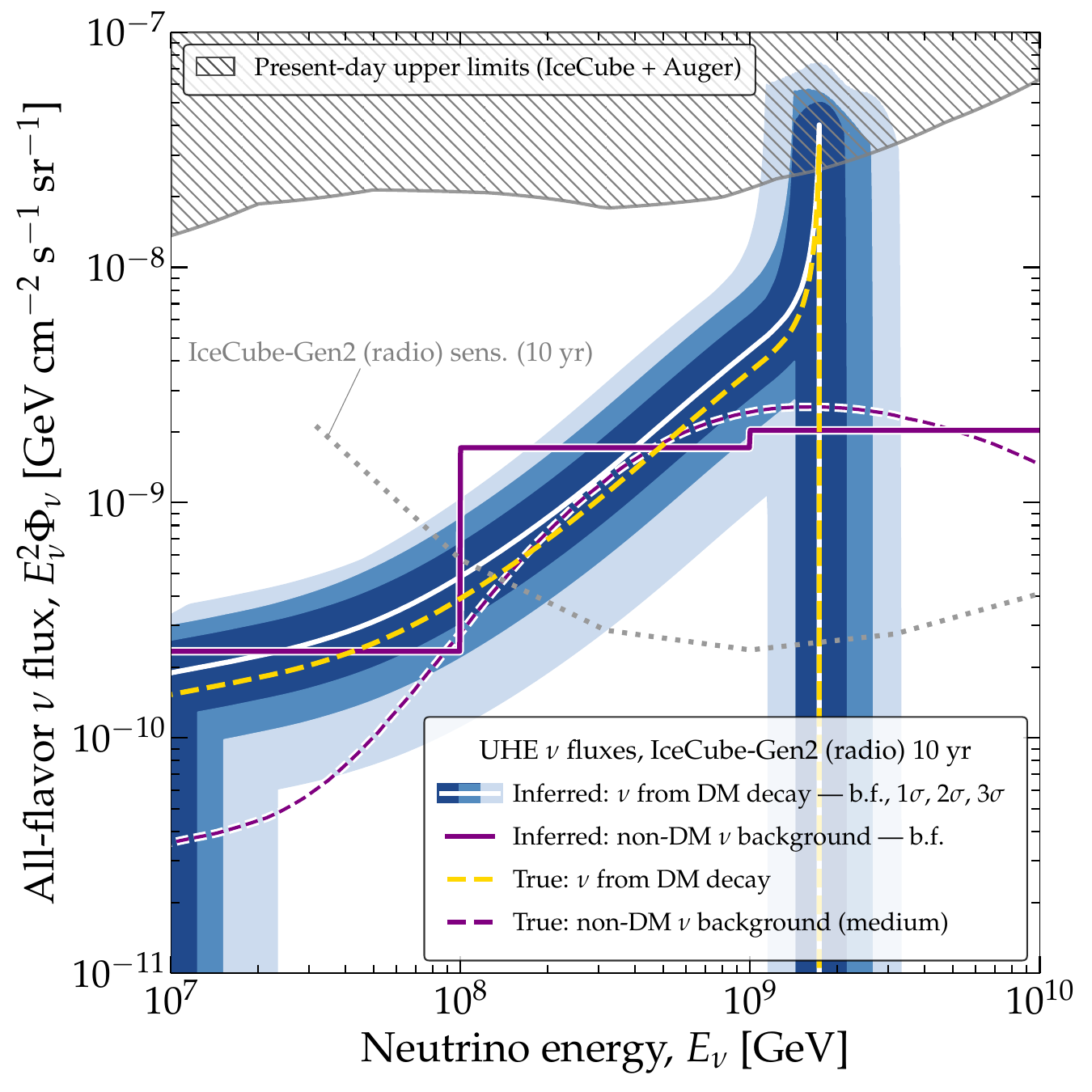}
 \caption{\textit{\textbf{Inferred flux of UHE neutrinos using the radio array of IceCube-Gen2.}}  The flux of neutrinos from DM decay, sky-averaged, and the isotropic background flux are shown separately.  The true values of the DM mass and lifetime are the same as used in \figu{reconstructed}.  The allowed band of neutrino flux from DM decay corresponds to the allowed regions of inferred DM mass and lifetime from \figu{reconstructed}.  The forecast assumes our medium-sized benchmark background flux of non-DM UHE neutrinos (Sec.~\ref{sec:fluxes_astro}), \ie, 10\% of the cosmogenic neutrino flux by Bergman \& van Vliet~\cite{Anker:2020lre}  (\figu{diffuse_fluxes}), or about 3 background events detected per year.  This figure assumes the NFW density profile for Galactic DM.  Existing upper limits on the flux of UHE neutrinos are from IceCube~\cite{IceCube:2018fhm} and Auger~\cite{PierreAuger:2019ens}.  The projected sensitivity of the radio array of IceCube-Gen2 is from \Refe~\cite{IceCube-Gen2:2021rkf}. \textit{Even using a coarse model for the UHE neutrino flux background (\equ{flux_bg_pw}), the flux of neutrinos from DM decay may be reconstructed with reasonable accuracy and precision.} See Sec.~\ref{sec:discovery_measuring} for details.}\label{fig:flux_reconstructed}
\end{figure}

Figure~\ref{fig:reconstructed} illustrates that, in the event of discovery of DM decay into UHE neutrinos, IceCube-Gen2 may infer the values of the DM mass and lifetime responsible for the discovered signal.  In \figu{reconstructed}, we show results obtained using our illustrative choice for the true DM and mass lifetime (see above).  To illustrate the influence of a non-DM isotropic background flux of UHE neutrinos, we compare results obtained assuming no background {\it vs.}~assuming our medium benchmark background (Sec.~\ref{sec:fluxes_astro}).

Our results confirm our expectation (Sec.~\ref{sec:discovery_measuring_stat}) that using the energy distribution of events grants us sensitivity to the DM parameters.  For the choice of DM mass and lifetime in \figu{reconstructed}, and similar ones, their values can be inferred with an accuracy of a factor of 2--3.  In the presence of the medium background, the accuracy degrades only slightly compared to the null-background case.  A larger background degrades the accuracy further, but does not preclude measurement; it may, however, reduce the precision with which the DM mass is inferred (see below).

While the accuracy on the DM mass is only weakly degraded by the presence of a background, the precision on it suffers more appreciably.  In the absence of a background, the best-fit value of the DM mass matches its true value.  In the presence of a medium-size background, its best-fit value is offset from the real one.  This stems from our choice of the three-piece background flux, \equ{flux_bg_pw}, to analyze projected event samples.  On the one hand, this flux prescription frees us from having to rely on specific theoretical predictions of the UHE astrophysical or cosmogenic neutrino flux.  On the other hand, because it is rather coarse---with decade-wide flux normalization constants---it fails to reproduce closely the shape of the background neutrino energy spectrum, \figu{diffuse_fluxes}.   Even so, the mismatch between the best-fit and true values is small; they are consistent within $1\sigma$.  Future analyses could mitigate this loss of precision by adopting a more finely binned version of the piecewise background flux.

\textbf{\textit{Thus, the DM mass and lifetime can be accurately inferred, even in the presence of a non-DM isotropic background flux of UHE neutrinos, by analyzing jointly the angular and energy distribution of events.}}  In doing so, there is essentially no degeneracy between the flux of UHE neutrinos from DM decay and the unknown background flux of UHE astrophysical and cosmogenic neutrinos, since in our procedure the former is determined almost exclusively by neutrinos from the GC, while the latter is determined by neutrinos from every direction.

Figure~\ref{fig:flux_reconstructed} shows the corresponding allowed regions of the sky-averaged diffuse flux of UHE neutrinos from DM.  The approximate factor-of-2 uncertainty on the DM lifetime translates into an uncertainty of similar size on the flux normalization.  The mismatch between the best-fit and true vales of the DM mass translates into a mismatch between the low-energy tails of secondary neutrinos from electroweak corrections in their corresponding fluxes.  Figure~\ref{fig:flux_reconstructed} shows also that our piecewise background flux model, \equ{flux_bg_pw}, is able to match the true background flux reasonably well, within the limitations of its coarse shape, except in the lowest energy bin, where the match is poor because due to the background flux dipping well below the sensitivity of the radio array of IceCube-Gen2, thus leading to low event rates.  There, results could be improved by a combined analyses of TeV--PeV and UHE neutrinos detected, respectively, by the optical and radio arrays of IceCube-Gen2~\cite{IceCube-Gen2:2020qha, vanSanten:2022wss}.


\section{Projected bounds on dark matter decay}
\label{sec:bounds}

\textit{The lifetime of heavy DM that decays into UHE neutrinos may be bound even in the presence of sizable non-DM neutrino backgrounds, by using the joint angular and energy distribution of detected events, in 10~years of exposure of the radio array of IceCube-Gen2 (\figu{bounds_medium_bg}).  Even when using the largest possible allowed background, the bounds on the DM lifetime remain competitive or better than present-day bounds (\figu{bounds}).}


\subsection{Overview}
\label{sec:bounds_overview}

Absent evidence for UHE neutrinos from DM decay, it may still be possible to place competitive bounds on the DM lifetime, even in the presence of a sizable non-DM isotropic UHE neutrino background flux. Like for the discovery of DM decay (Sec.~\ref{sec:discovery}), below we gear our results for radio-detection at IceCube-Gen2, but our methods and, broadly stated, our conclusions are applicable to next-generation UHE neutrino telescopes in general.

References~\cite{Guepin:2021ljb, Chianese:2021jke, Arguelles:2022nbl} reported projected bounds for DM decay into UHE neutrinos, including via their radio-detection at IceCube-Gen2.  We improve on those in two ways.  First, we use a significantly more detailed calculation of event rates in IceCube-Gen2, based on state-of-the-art simulations of neutrino propagation, interaction, and radio-detection (Sec.~\ref{sec:detection}).  Like for the discovery of DM decay, this is key to generating reliable angular and energy event distributions, which our analysis uses to discriminate against the background.  Second, unlike previous works, we forecast bounds in the presence of a sizable non-DM isotropic UHE neutrino background flux.  Reference~\cite{Chianese:2021jke} did consider the presence of a potential background, but discriminated against it simply by counting only neutrinos with energy smaller than $m_\mathrm{DM} / 2$.  In contrast, we use a full-fledged angular and energy analysis to produce our bounds.


\subsection{Statistical methods}
\label{sec:bounds_stat}

Unlike our earlier analyses to discover DM decay and infer the DM mass and lifetime (Sec.~\ref{sec:discovery}), to place bounds on DM decay we assume that the {\it true}, observed event distributions are due solely to the non-DM isotropic background flux of UHE neutrinos, and we contrast {\it test} event distributions expected from DM decay against it.  For the non-DM background flux, we use our two medium and large benchmark fluxes (Sec.~\ref{sec:fluxes_astro}), based off of the cosmogenic neutrino flux by Bergman \& van Vliet~\cite{Anker:2020lre}.  Our main conclusions hold for other choices of background, and we point out below what features of our results are due to our specific benchmark choices.  

Further, unlike when computing discovery prospects (Sec.~\ref{sec:discovery_prospects}), when setting bounds below we use not only the angular distribution of events, but also their energy distribution.  We demand that the energy distribution of events lacks features that are characteristic of the energy spectrum of neutrinos from DM decay, \ie, a clustering of events with similar energies that reflects an underlying bump-like shape of the spectrum (Sec.~\ref{sec:fluxes_dm}).  Admittedly, this is a broad, conservative criterion: it discriminates against the flux of neutrinos from DM decay, but also against any astrophysical or cosmogenic flux that has a bump-like feature in its spectrum, of which there are many proposals, including our benchmark background fluxes; see, \eg, Fig.~2 in \Refe~\cite{Valera:2022wmu}.   

For a given choice of the non-DM neutrino background, we compute the true, observed event distribution, 
\begin{equation}
 \frac{dN_\nu^{\rm true}}{dE^\mathrm{rec}_\mathrm{sh}d\Omega^\mathrm{rec}}
 =
 \frac{dN_\nu^{\rm bg}}{dE^\mathrm{rec}_\mathrm{sh}d\Omega^\mathrm{rec}} \;,
\end{equation}
using the methods from Sec.~\ref{sec:detection_rates}.  From it, we sample random realizations of the observed event sample, consisting of a random number $N_{\rm evts}$ of events, each with reconstructed energy and direction, $E_{{\rm sh}, i}^{\rm rec}$ and $\Omega_i^{\rm rec}$.

Then, for a choice of the DM mass $m_\mathrm{DM}$ and lifetime $\tau_\mathrm{DM}$, we compare the true event rate {\it vs.}~the test event rate expected from DM decay, $dN_\nu^{\rm test}(\boldsymbol{\vartheta}^\prime) / dE^\mathrm{rec}_\mathrm{sh} d\Omega^\mathrm{rec}$, given by \equ{test_event_rate_measure} evaluated at test parameters $\boldsymbol{\vartheta}^\prime \equiv (m_{\rm DM}^\prime = m_{\rm DM}, \tau_{\rm DM}^\prime = \tau_{\rm DM}, 
\mathcal{N}_{\Phi, 1}^\prime, \mathcal{N}_{\Phi, 2}^\prime, \mathcal{N}_{\Phi, 3}^\prime)$.  This test event rate is computed using the same piecewise background UHE neutrino spectrum, \equ{flux_bg_pw}, that we used to infer the DM mass and lifetime (Sec.~\ref{sec:discovery_measuring}).

Like when computing discovery forecasts, we compare the true (\ie, background-only) and test (\ie, background plus DM) hypotheses via an unbinned likelihood, given by \equ{likelihood_mea}, \ie,
\begin{eqnarray}
 \label{equ:likelihood_full}
 \mathcal{L}_{\rm full}(\boldsymbol{\vartheta}^\prime, 
 \left\{ E_{{\rm sh}, i}^{\rm rec}, \Omega_i^{\rm rec} \right\})
 &=&
 e^{-N_\nu^{\rm test}(\boldsymbol{\vartheta}^\prime)}
 \\
 && \nonumber
 \times
 \prod_{i=1}^{N_\mathrm{evts}}
 \left.
 \frac{dN_\nu^{\rm test}(\boldsymbol{\vartheta}^\prime)}{dE^\mathrm{rec}_\mathrm{sh}d\Omega^\mathrm{rec}}
 \right\vert_{E_{{\rm sh}, i}^{\rm rec}, \Omega_i^{\rm rec}}
 \;,
\end{eqnarray}
which relies on the full available information on the events, \ie, their joint angular and energy distribution.

In addition, to highlight the role of the angular and energy information in placing bounds, we compute separately analyses that use limited information.  An analysis that relies only on angular information uses a likelihood where the event rate is integrated across all reconstructed energies from $10^7$~GeV to $10^{10}$~GeV, \ie,
\begin{eqnarray}
 \label{equ:likelihood_ang}
 \mathcal{L}_{\rm ang}(\boldsymbol{\vartheta}^\prime, 
 \left\{ \Omega_i^{\rm rec} \right\})
 &=&
 e^{-N_\nu^{\rm test}(\boldsymbol{\vartheta}^\prime)}
 \\
 && \nonumber
 \times
 \prod_{i=1}^{N_\mathrm{evts}}
 \left.
 \left(
 \int dE_{\rm sh}^{\rm rec}
 \frac{dN_\nu^{\rm test}(\boldsymbol{\vartheta}^\prime)}{dE^\mathrm{rec}_\mathrm{sh}d\Omega^\mathrm{rec}}
 \right)
 \right\vert_{\Omega_i^{\rm rec}}
 \;.
\end{eqnarray}
An analysis that relies only on energy information uses a likelihood where the event rate is all-sky, \ie,
\begin{eqnarray}
 \label{equ:likelihood_en}
 \mathcal{L}_{\rm en}(\boldsymbol{\vartheta}^\prime, 
 \left\{ E_{{\rm sh}, i}^{\rm rec} \right\})
 &=&
 e^{-N_\nu^{\rm test}(\boldsymbol{\vartheta}^\prime)}
 \\
 && \nonumber
 \times
 \prod_{i=1}^{N_\mathrm{evts}}
 \left.
 \left(
 \int d\Omega^{\rm rec}
 \frac{dN_\nu^{\rm test}(\boldsymbol{\vartheta}^\prime)}{dE^\mathrm{rec}_\mathrm{sh}d\Omega^\mathrm{rec}}
 \right)
 \right\vert_{E_{{\rm sh}, i}^{\rm rec}}
 \;.
\end{eqnarray}
And, finally, an analysis that relies only on the all-sky number of events of all energies uses
\begin{equation}
 \label{equ:likelihood_count}
 \mathcal{L}_{\rm count}(\boldsymbol{\vartheta}^\prime) 
 =
 e^{-N_\nu^{\rm test}(\boldsymbol{\vartheta}^\prime)}
 \left[
 N_\nu^{\rm test}(\boldsymbol{\vartheta}^\prime)
 \right]^{N_{\rm evts}} \;.
\end{equation}

Like for the DM discovery prospects before, we compute projected bounds on the DM mass and lifetime using an Asimov event sample.  In analogy to \equ{likelihood_dm_avg}, we average the above likelihood functions over all possible random realizations of the observed events, sampled from the underlying event distribution due to the non-DM background, $dN_\nu^{\rm true} / dE^\mathrm{rec}_\mathrm{sh} d\Omega^\mathrm{rec}$. This yields the average functions $\langle \ln \mathcal{L}_{\rm full} \rangle$, $\langle \ln \mathcal{L}_{\rm ang} \rangle$, $\langle \ln \mathcal{L}_{\rm en} \rangle$, and $\langle \ln \mathcal{L}_{\rm count} \rangle$.  In analogy to \equ{likelihood_avg_bg}, we define likelihood functions computed under the background-only  hypothesis, \eg,
\begin{eqnarray}
 &&
 \langle 
 \ln \mathcal{L}_{\rm full, bg} 
 \left(
 \mathcal{N}_{\Phi, 1}^\prime, 
 \mathcal{N}_{\Phi, 2}^\prime, 
 \mathcal{N}_{\Phi, 3}^\prime
 \right)
 \rangle
 \\
 && \nonumber
 =
 \lim_{\tau_{\rm DM}^\prime \to \infty}
 \langle 
 \ln \mathcal{L}_{\rm full} 
 \left(
 m_{\rm DM}^\prime,
 \tau_{\rm DM}^\prime,
 \mathcal{N}_{\Phi, 1}^\prime, 
 \mathcal{N}_{\Phi, 2}^\prime, 
 \mathcal{N}_{\Phi, 3}^\prime
 \right)
 \rangle \;,
\end{eqnarray}
where the right-hand side no longer depends on the DM mass and lifetime.  Similar expressions apply for the other likelihood functions, \ie, $\langle \mathcal{L}_{\rm ang, bg} \rangle$, $\langle \mathcal{L}_{\rm en, bg} \rangle$, $\langle \mathcal{L}_{\rm count, bg} \rangle$.

To place bounds, we follow \Refe~\cite{Cowan:2010js} and define a test statistic that compares the true hypothesis---that there is no DM neutrino flux---and test hypothesis---that there is a DM neutrino flux with parameters $m_\mathrm{DM}$ and $\tau_\mathrm{DM}$.  \Eg, for the full analysis,
\begin{widetext}
\begin{eqnarray}
 \langle
 \Lambda_{\rm full} (m_{\rm DM}, \tau_{\rm DM})
 \rangle
 &=&
 -2~
 {\rm min}_{
 \mathcal{N}_{\Phi, 1}^\prime, 
 \mathcal{N}_{\Phi, 2}^\prime, 
 \mathcal{N}_{\Phi, 3}^\prime}
 \left[
 \langle 
 \ln \mathcal{L}_{\rm full} 
 \left(
 m_{\rm DM},
 \tau_{\rm DM},
 \mathcal{N}_{\Phi, 1}^\prime, 
 \mathcal{N}_{\Phi, 2}^\prime, 
 \mathcal{N}_{\Phi, 3}^\prime
 \right)
 \rangle
 -
 \langle 
 \ln \mathcal{L}_{\rm full,bg} 
 \left(
 \mathcal{N}_{\Phi, 1}^\prime, 
 \mathcal{N}_{\Phi, 2}^\prime, 
 \mathcal{N}_{\Phi, 3}^\prime
 \right)
 \rangle
 \right]
 \nonumber \\
 && 
 \times ~
 \Theta\left[\hat{\tau}_\mathrm{DM}-\tau_\mathrm{DM}\right] \;,
\end{eqnarray}
\end{widetext}
where $\Theta$ is the Heaviside function and $\hat{\tau}_\mathrm{DM}(m_\mathrm{DM})$ is the value of the DM lifetime that, for a fixed value of the DM mass, $m_\mathrm{DM}$, maximizes the likelihood function $\langle \ln \mathcal{L}_{\rm full} \rangle$.  Similar expressions apply for the other analyses, \ie, $\langle \Lambda_{\rm ang} \rangle$, $\langle \Lambda_{\rm en} \rangle$, $\langle \Lambda_{\rm count} \rangle$.  With this definition, under the null hypothesis where neutrinos from DM decay exist, the test statistic should be distributed according to a half-$\chi^2$ distribution with one degree of freedom.  Hence, below, we place limits on the DM lifetime at the $2\sigma$~C.L. when $\langle \Lambda_{\rm full} \rangle > 2.7$, and similarly for the other analyses.


\subsection{Results}
\label{sec:bounds_results}

\begin{figure}
 \includegraphics[width=\columnwidth]{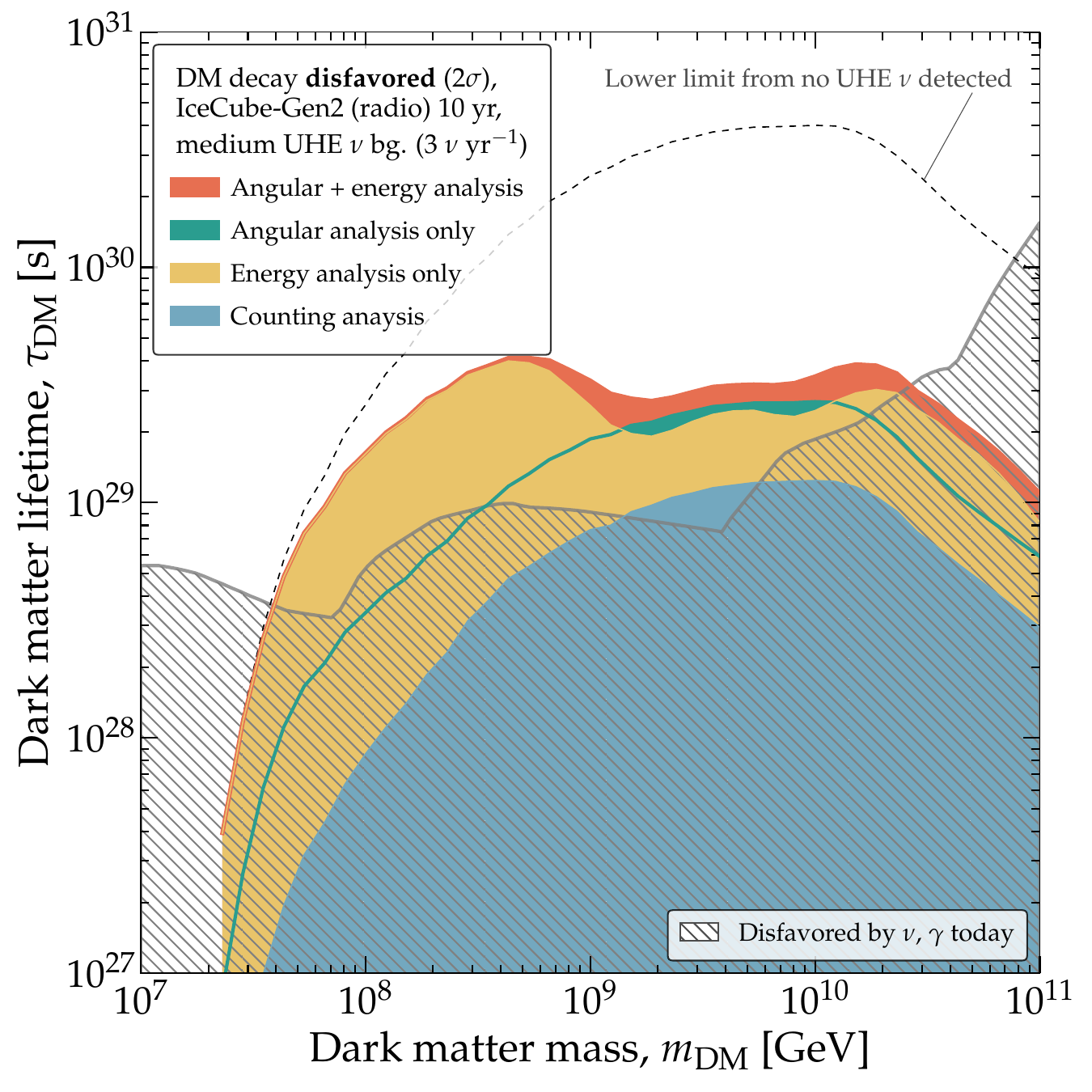}
 \caption{\textbf{\textit{Projected lower limits on the DM mass and lifetime from a search for UHE neutrinos in the radio array of IceCube-Gen2.}}  The limits are placed in the absence of detection of neutrinos from DM decay, assuming a medium-sized benchmark non-DM isotropic background flux of UHE neutrinos, \ie, 10\% of the cosmogenic neutrino flux by Bergman \& van Vliet~\cite{Anker:2020lre}; see \figu{diffuse_fluxes}.  We show bounds obtained using the joint angular and energy distribution of events---our main results---only the angular or only the energy distribution, and only counting the all-sky rate of events of all energies.  The bounds obtained using energy information dip between $10^9$ and $10^{10}$~GeV, where our benchmark background neutrino energy spectrum peaks and may mimic the spectrum of neutrinos from DM decay.  We adopt baseline choices of the detector angular and (logarithmic) energy resolution, $\sigma_{\Omega} = 3^\circ$ and $\sigma_{\epsilon} = 0.1$; see Sec.~\ref{sec:detection_simulation}.  For comparison, we show limits obtained if no UHE neutrinos are detected at $2\sigma$.  This figure assumes the NFW density profile for Galactic DM.  Existing lower limits on the DM lifetime are the same as in \figu{summary}.  \textit{Even in the presence of a sizable non-DM isotropic background flux of UHE neutrinos, using the angular and energy distributions of events keeps the projected lower limits on DM lifetime comparable to, or better than present-day ones.}   See \figu{bounds_large_bg} for results obtained using a large UHE neutrino background.  See Sec.~\ref{sec:bounds} for details.
 }\label{fig:bounds_medium_bg}
\end{figure}

\begin{figure}
 \includegraphics[width=\columnwidth]{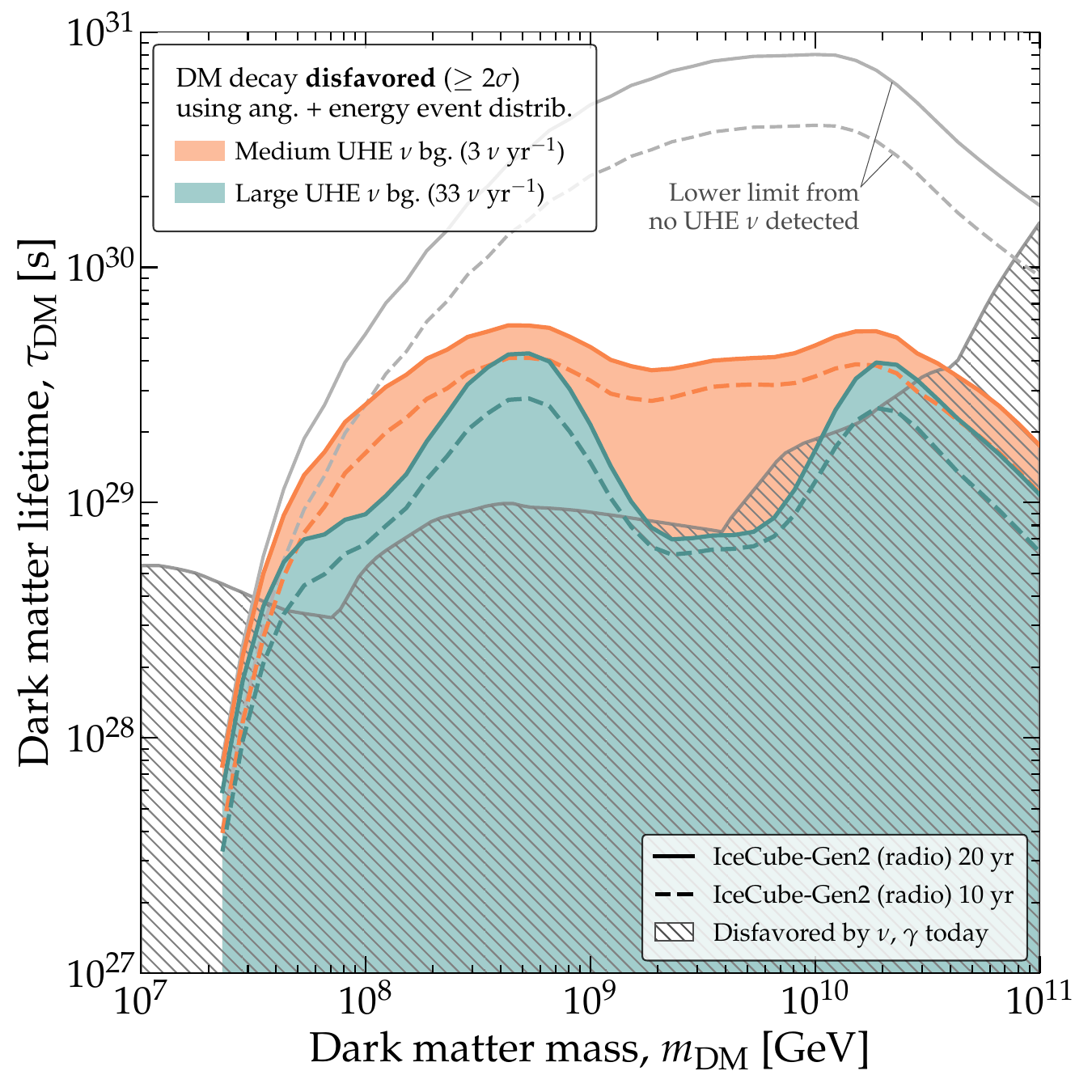}
 \caption{\textbf{\textit{Projected lower limits on the DM lifetime for different choices of non-DM UHE neutrino background.}}  Same as \figu{bounds_medium_bg}, but showing only results obtained from the joint angular and energy analysis.  We compare results obtained using our medium and large benchmark background isotropic fluxes of non-DM neutrinos (\figu{diffuse_fluxes}), and using 10~years and 20~years of detector exposure.  The limit on DM lifetime is degraded for values of the DM mass that yield neutrino spectra that peak at energies where our benchmark background neutrino spectrum also peaks, \ie, between $10^9$ and $10^{10}$~GeV.  This figure assumes the NFW density profile for Galactic DM. \textit{The presence of a sizable non-DM isotropic background flux of UHE neutrinos will degrade the lower limits that can be placed on the DM lifetime, but only within a window of values of the DM mass.}  See Sec.~\ref{sec:bounds} for details.
 }\label{fig:bounds}
\end{figure}

Figure~\ref{fig:bounds_medium_bg} shows the resulting projected bounds on the DM lifetime, obtained by adopting our benchmark medium non-DM UHE neutrino background.  We extract two main observations from it. 

First, \figu{bounds_medium_bg} shows that the existence of a sizable non-DM neutrino background appreciably weakens the bounds, compared to those obtained by plainly demanding that no UHE neutrino is detected, which are representative of most previous analyses in the literature.  (In \figu{bounds_medium_bg}, the null-detection curve also corresponds to a value of the test statistic of 2.7, which implies a mean number of detected events of 1.35.)  Blatantly, when using only a counting analysis, the bounds that we obtain are up to 40 times weaker than bounds obtained from demanding that no neutrino is detected.  In reality, how much the bounds are weakened will depend on the actual size and shape of the non-DM background.  Still, our results serve as a reminder that projected bounds on the DM lifetime reported in the literature may be optimistic.

Second, \figu{bounds_medium_bg} shows that using the angular and energy event distributions mitigates how much the bounds are weakened.  Depending on the DM mass, they improve the bounds compared to the counting analysis by a factor of 2--10.  This holds even when adopting our benchmark large background instead; see \figu{bounds_large_bg}. Overall, using the angular and energy  distributions allows projected bounds to remain competitive with present-day ones.

The angular and energy information complement each other.  Using the energy distribution strongly improves the bounds at low and high DM masses.  In the intermediate region, between $10^9$~GeV and $10^{10}$~GeV, bounds that use energy information weaken because this is where our benchmark background neutrino spectrum peaks (\figu{diffuse_fluxes}) and where it may be misconstrued as being due to DM decay; see the discussion in Sec.~\ref{sec:bounds_stat}.  This is counteracted by using angular information: in the $10^9$--$10^{10}$~GeV range, where the detector response is largest (see Fig.~13 in \Refe~\cite{Valera:2022ylt}), the isotropic neutrino background induces a number of events large enough for the analysis to reject an excess towards the GC from DM decay.  Figure~\ref{fig:bounds_medium_bg} shows the result of this interplay for our particular choice of non-DM background; in reality, the specifics will depend on the actual size and shape of the background.

Figure~\ref{fig:bounds} shows that using our large benchmark non-DM UHE neutrino background instead---ten times larger than the medium one---weakens the bounds by only a factor of roughly 2, except in the range $10^9$--$10^{10}$~GeV, where the bounds weaken by a factor of up to 6, but even so remain roughly competitive with present-day ones.  This represents promising prospects:
\textbf{\textit{the background flux we use here~\cite{Anker:2020lre} is as large as allowed by the present-day IceCube~\cite{IceCube:2018fhm} and Auger~\cite{PierreAuger:2019ens} upper limits. Yet, even with this aggressive choice of background, our projected bounds in \figu{bounds} remain comparable or better than present-day ones.}}

\begin{figure}
 \includegraphics[width=\columnwidth]{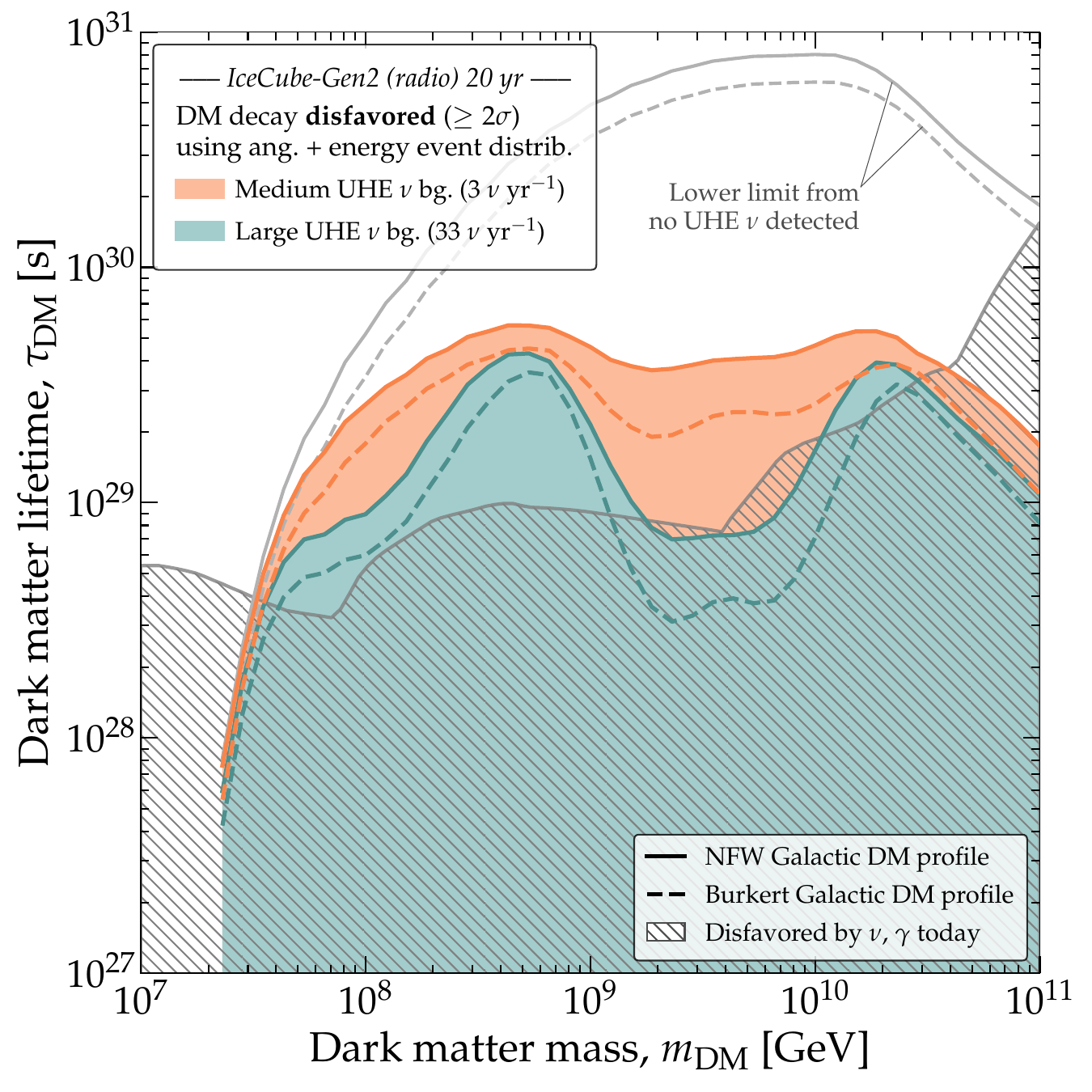}
 \caption{\textit{\textbf{Projected lower limits on the DM lifetime for NFW vs.~Burkert Galactic DM profiles.}}  Same as \figu{bounds}, but comparing results obtained using the NFW {\it vs.}~Burkert Galactic DM density profiles (see \figu{dm_profiles}).  In this figure, results are for discovery at $\geq 2 \sigma$ after 20 years; see \figu{bounds_burkert} for more results using the Burkert profile.  \textit{If the Galactic DM profile is ``puffy'', like the Burkert profile, bounds weaken by a factor of roughly 2, compared to a ``cuspy'' DM profile, like NFW.}  See Sec.~\ref{sec:bounds} for details.}\label{fig:bounds_nfw_vs_burkert}
\end{figure}

Figure~\ref{fig:bounds_nfw_vs_burkert} shows that, naturally, the bounds degrade when using the Burkert Galactic DM profile instead of the NFW profile.  Because the Burkert profile is puffier, the bounds derived from the analysis of the angular distribution of events are weakened, so most of the limit-setting power comes from the energy distribution of events instead; see also \figu{bounds_burkert}.  Still, the energy analysis sets bounds using the Burkert profile that are only a factor-of-2 worse than using the NFW profile.  Thus, given our extant imperfect knowledge of the Galactic DM profile---in particular, given the possibility of a puffy DM profile---leveraging the interplay between energy and angular information is key.

\section{Summary and outlook}
\label{sec:summary}

In the next decade, ultra-high-energy (UHE) neutrino telescopes, presently in planning, will deliver a new way to look for heavy dark matter (DM), with masses in excess of $10^7$~GeV, via its decay into UHE neutrinos, with energies in excess of $10^7$~GeV.  To properly harness this potential, it is critical to disentangle the signatures of UHE neutrinos of DM decay origin from the signatures of UHE neutrinos of astrophysical and cosmogenic origin---long-sought but still undiscovered---that act as a background to DM searches.  Failure to do so may incur in steep misrepresentation when claiming discovery of DM decay, inferring the DM mass and lifetime in the event of discovery, or setting bounds on the DM mass and lifetime otherwise.  The task is complicated by the fact that the size and shape of the non-DM neutrino background is unknown, that the number of detected events may be small, and that the direction- and energy-measurement capabilities of the detectors are limited.

Even so, we have shown, by means of detailed forecasts, that these obstacles are surmountable.  Key to that is to examine the energy and angular distributions of the detected UHE neutrinos.  They grant us access to the essential differences between the diffuse neutrino fluxes from DM decay and from the non-DM background: in energy, the former is concentrated around the DM mass, while the latter is more spread out (\figu{diffuse_fluxes}), and, in direction, the former is concentrated around the Galactic Center (GC)---where DM is abundant---while the latter is isotropic (\figu{sky_map_flux}).  In our forecasts, we look for these differences in projected observations.

We have geared our forecasts to the radio-detection of UHE neutrinos in the envisioned IceCube-Gen2 neutrino telescope, which we simulate using state-of-the-art methods, including experimental nuance that dull the above differences between the fluxes.  Our findings are promising: these differences survive an analysis under realistic experimental conditions (Figs.~\ref{fig:diff_event_rate}, \ref{fig:sky_map_evrate}).  \textbf{\textit{Therefore, while the existence of a non-DM UHE neutrino background, even the largest presently allowed, weakens claims of discovery of DM decay or bounds on it, it does not necessarily preclude them}}.  Still, the limit-setting potential and, particularly, the discovery potential, are contingent to the Galactic DM profile peaking markedly towards the GC (Figs.~\ref{fig:discovery_prospects_nfw_vs_burkert} and \ref{fig:bounds_nfw_vs_burkert}).

Regarding the discovery of DM decay, we have shown that DM with mass between $10^8$~GeV and $10^{10}$~GeV and lifetime of roughly $10^{29}$~s should be discoverable after 10~years of operation of the radio array of IceCube-Gen2, even in the presence of a medium-sized non-DM neutrino background that yields about 3 events per year (\figu{discovery_prospects}).  This is conservatively achieved using only the angular distribution of detected events.  Under a larger background, of about 33 events per year, discovery becomes unfeasible in the face of existing bounds on the DM lifetime.  Our discovery forecasts depend only mildly on the shape of the energy spectrum of the non-DM neutrino background---whose size we let float in our analyses---and depend mainly on the total number of events.  

In the event of discovery, the DM mass and lifetime could be measured with reasonable accuracy and precision (\figu{reconstructed}), depending on their true values, and the flux of UHE neutrinos from DM decay could be similarly inferred (\figu{flux_reconstructed}).  Importantly, this result is robust: when inferring the values of the DM parameters---and also when setting bounds on them (see below)---we analyze the simulated event samples \textit{without} assuming knowledge of the shape and size of the energy spectrum of the non-DM neutrino background.

If discovery is not possible, we will be able to place lower limits on the DM lifetime, which we forecast.  Using the joint angular and energy distribution of events allows us to constrain the presence of bump-like energy spectra and excesses towards the Galactic Center even under challenging, but plausible scenarios where the energy spectrum of the non-DM neutrino background is medium-sized and whose bump-like energy spectrum inconveniently resembles that of neutrinos from DM decay (\figu{bounds_medium_bg}).  Even in the presence of a large bump-like non-DM background, using the energy distribution of events safeguards the limits on the DM lifetime for DM masses sufficiently far from the peak of the non-DM flux (\figu{bounds}).  Overall, we forecast lower limits on the DM lifetime that are comparable to, or better than, existing limits from gamma rays and TeV--PeV neutrinos.

While our forecasts are geared to the detection of UHE neutrinos in the radio array of IceCube-Gen2, our conclusions apply generally to planned neutrino telescopes of comparable size, radio-based and otherwise, and our methods can be readily adapted to them.  In particular, detectors with an envisioned high angular resolution, like GRAND, could be better at discovering or discriminating against an excess of events towards the GC~\cite{Guepin:2021ljb}.

It seems that, fortunately, the potential of next-generation UHE neutrino telescopes to probe heavy DM decay may be safeguarded against sizable unknown neutrino backgrounds.


\section*{Acknowledgements}

We thank Marco Chianese and Christian Glaser for useful discussions and Chengchao Yuan for comments on the manuscript.  DF, MB, and VBV are supported by the {\sc Villum Fonden} under project no.~29388. This work used resources provided by the High Performance Computing Center at the University of Copenhagen. This project has received funding from the European Union’s Horizon 2020 research and innovation program under the Marie Sklodowska-Curie Grant Agreement No.~847523 ‘INTERACTIONS’.


\appendix


\section{Additional figures}

\renewcommand{\theequation}{A\arabic{equation}}
\renewcommand{\thefigure}{A\arabic{figure}}
\setcounter{figure}{0} 

\begin{figure}[b!]
 \includegraphics[width=\columnwidth]{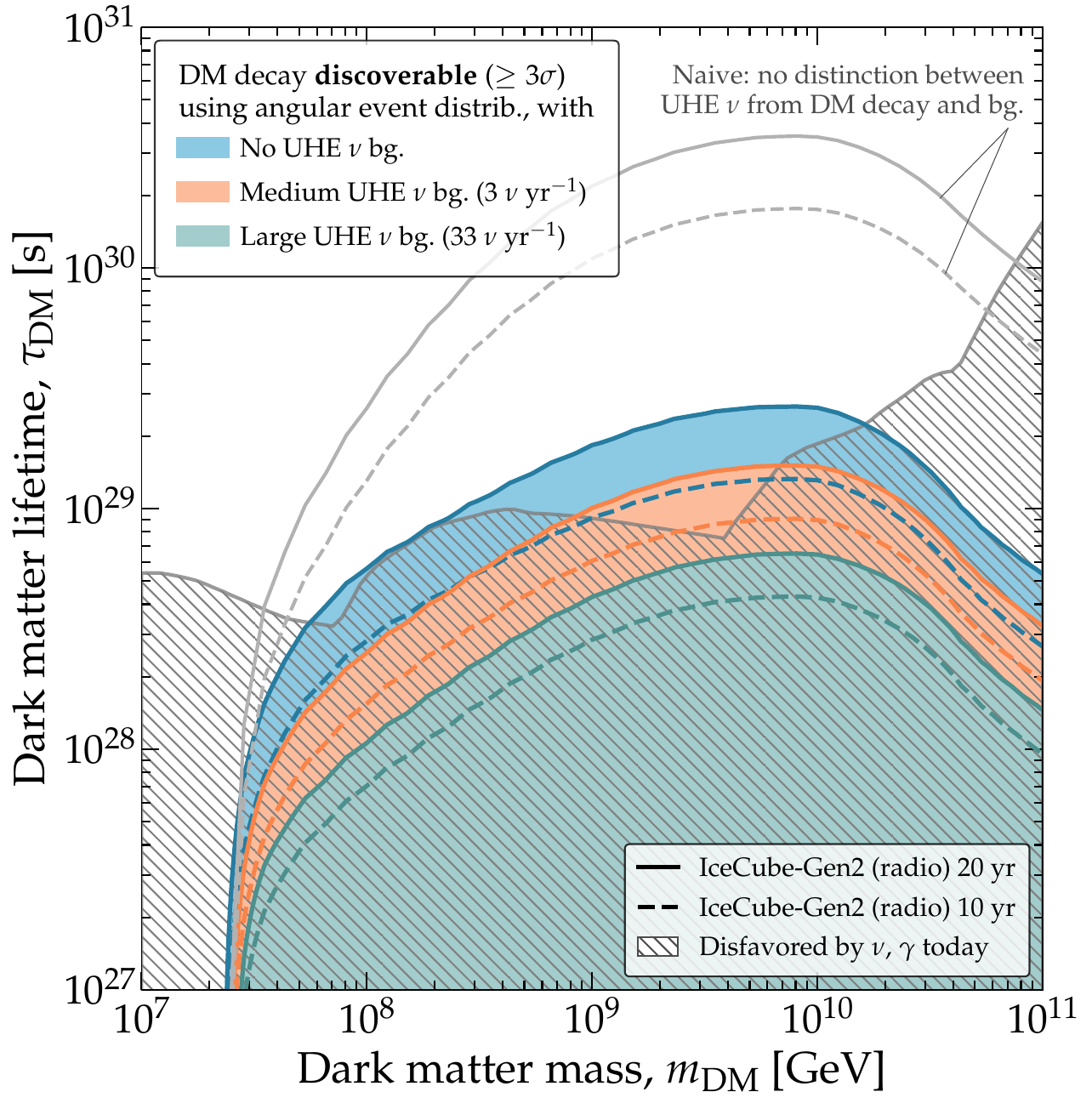}
 \caption{\textit{\textbf{Discovery prospects for UHE neutrinos from DM decay, at $\geq 3\sigma$.}} Compare to \figu{discovery_prospects}, which is for $\geq 2 \sigma$.}\label{fig:discovery_prospects_3sigma}
\end{figure}

\begin{figure}[b!]
 \includegraphics[width=\columnwidth]{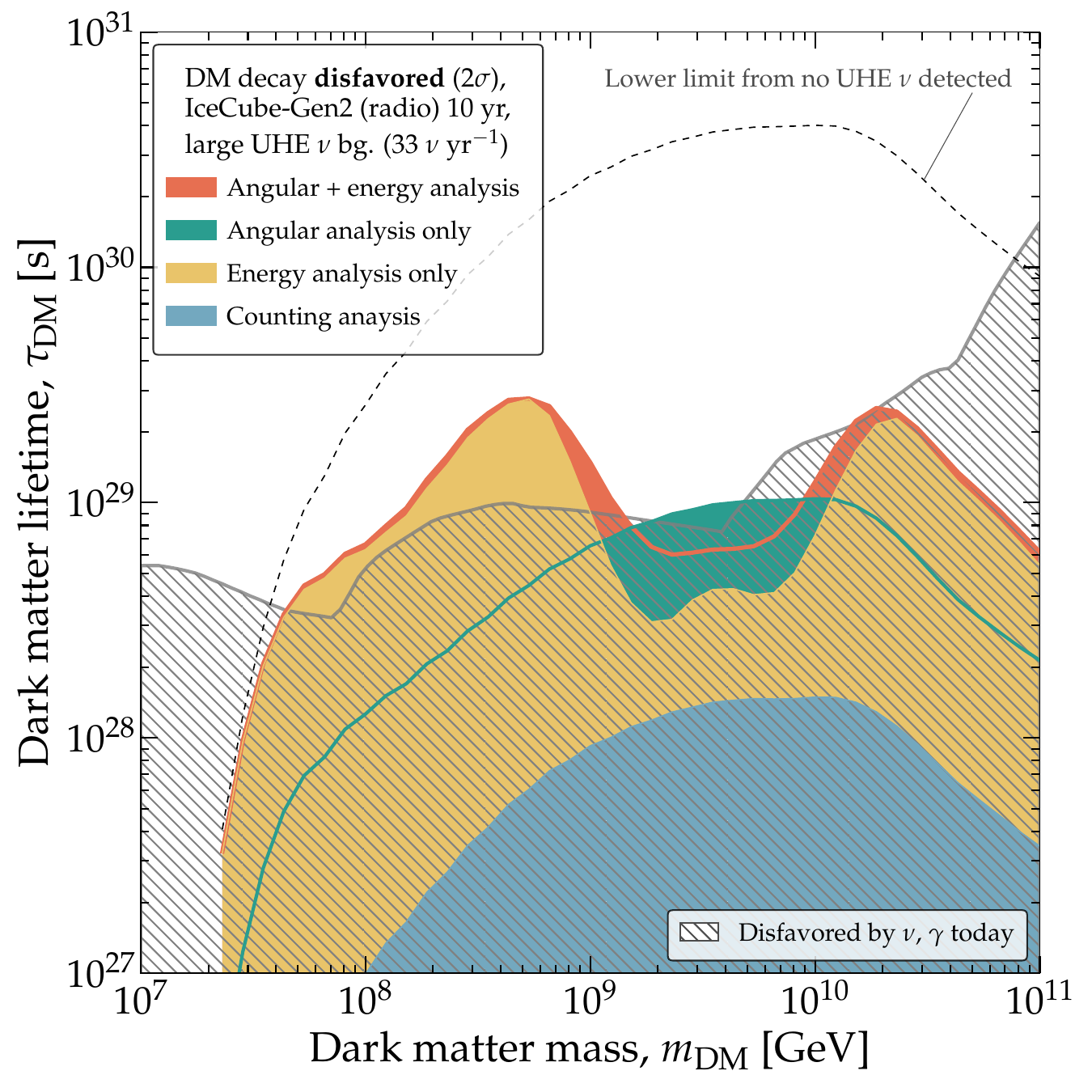}
 \caption{\textbf{\textit{Projected lower limits on the DM lifetime, under a large background of non-DM UHE neutrinos.}}  Same as \figu{bounds_large_bg}, but assuming our large benchmark non-DM isotropic background flux of UHE neutrinos, \ie, the cosmogenic neutrino flux by Bergman \& van Vliet~\cite{Anker:2020lre}.  See Sec.~\ref{sec:bounds} for details.}\label{fig:bounds_large_bg}
\end{figure}

\begin{figure*}[t!]
 \includegraphics[width=0.49\textwidth]{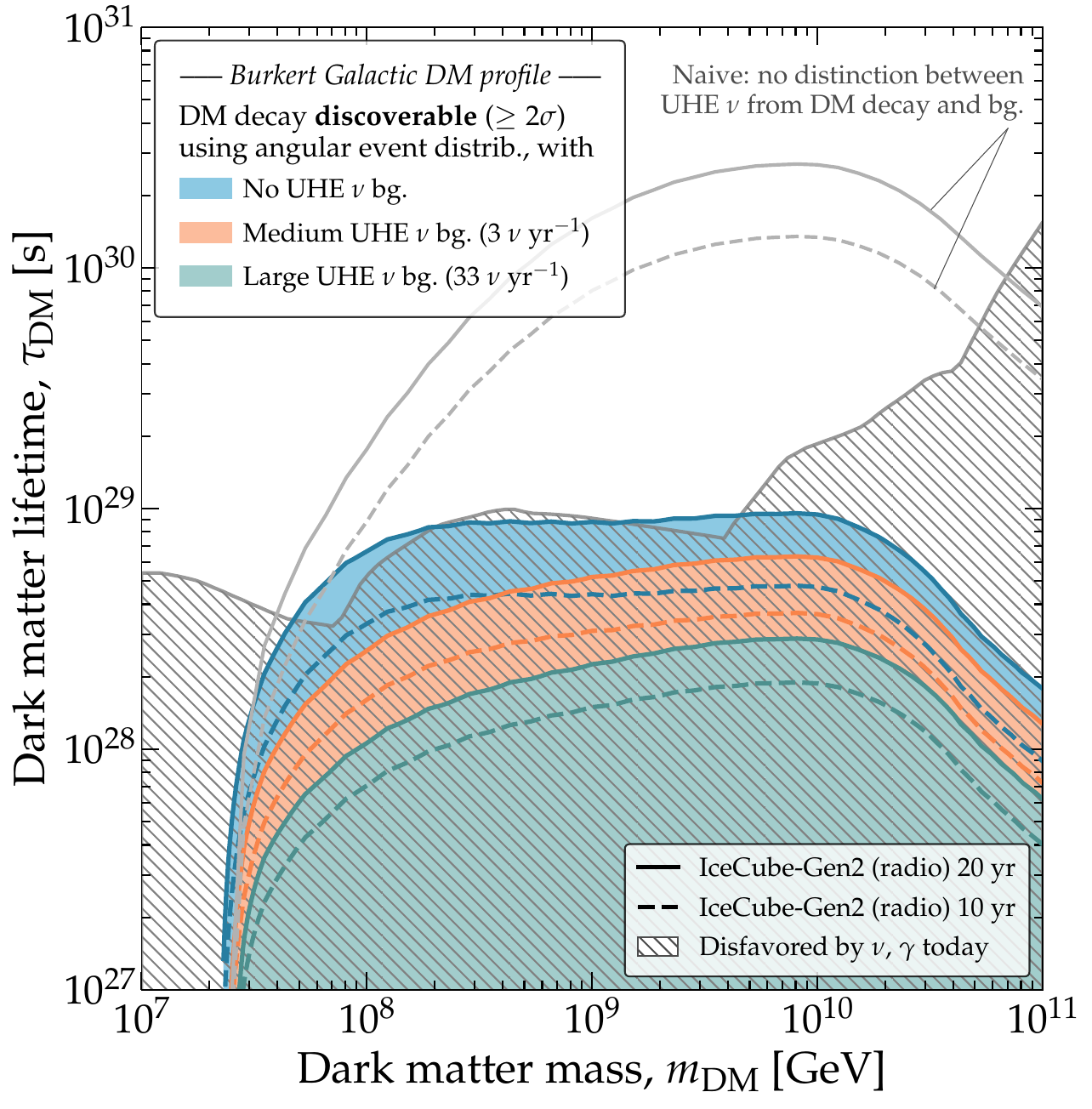}
 \includegraphics[width=0.49\textwidth]{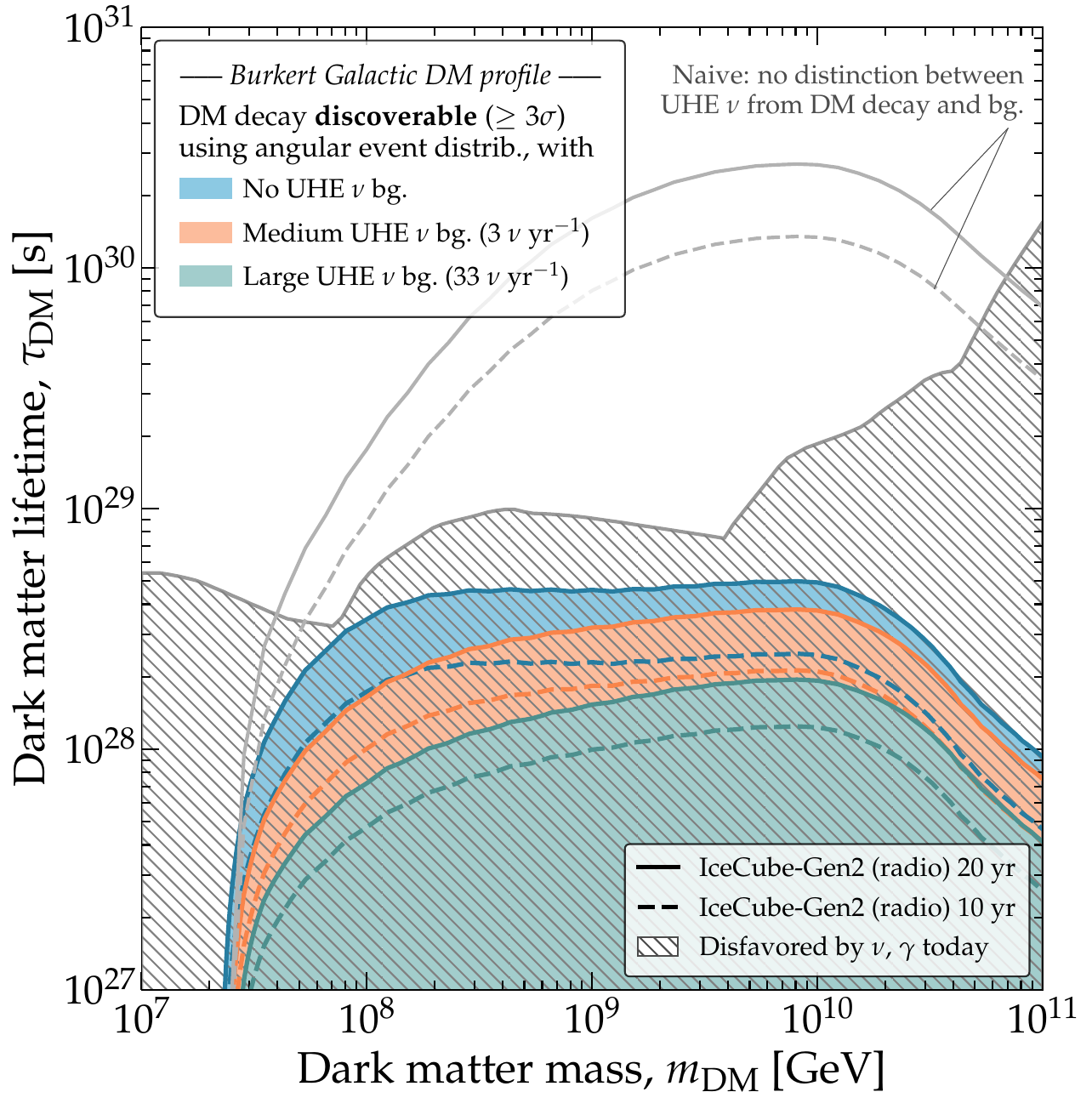}
 \caption{\textit{\textbf{Discovery prospects for UHE neutrinos from DM decay, using a Burkert Galactic DM profile.}} {\it Left:} For discovery at $\geq 2 \sigma$, to be compared with \figu{discovery_prospects}.  {\it Right:} For discovery at $\geq 3 \sigma$, to be compared with \figu{discovery_prospects_3sigma}.}\label{fig:discovery_prospects_burkert}
\end{figure*}

\begin{figure*}[t!]
 \includegraphics[width=0.49\textwidth]{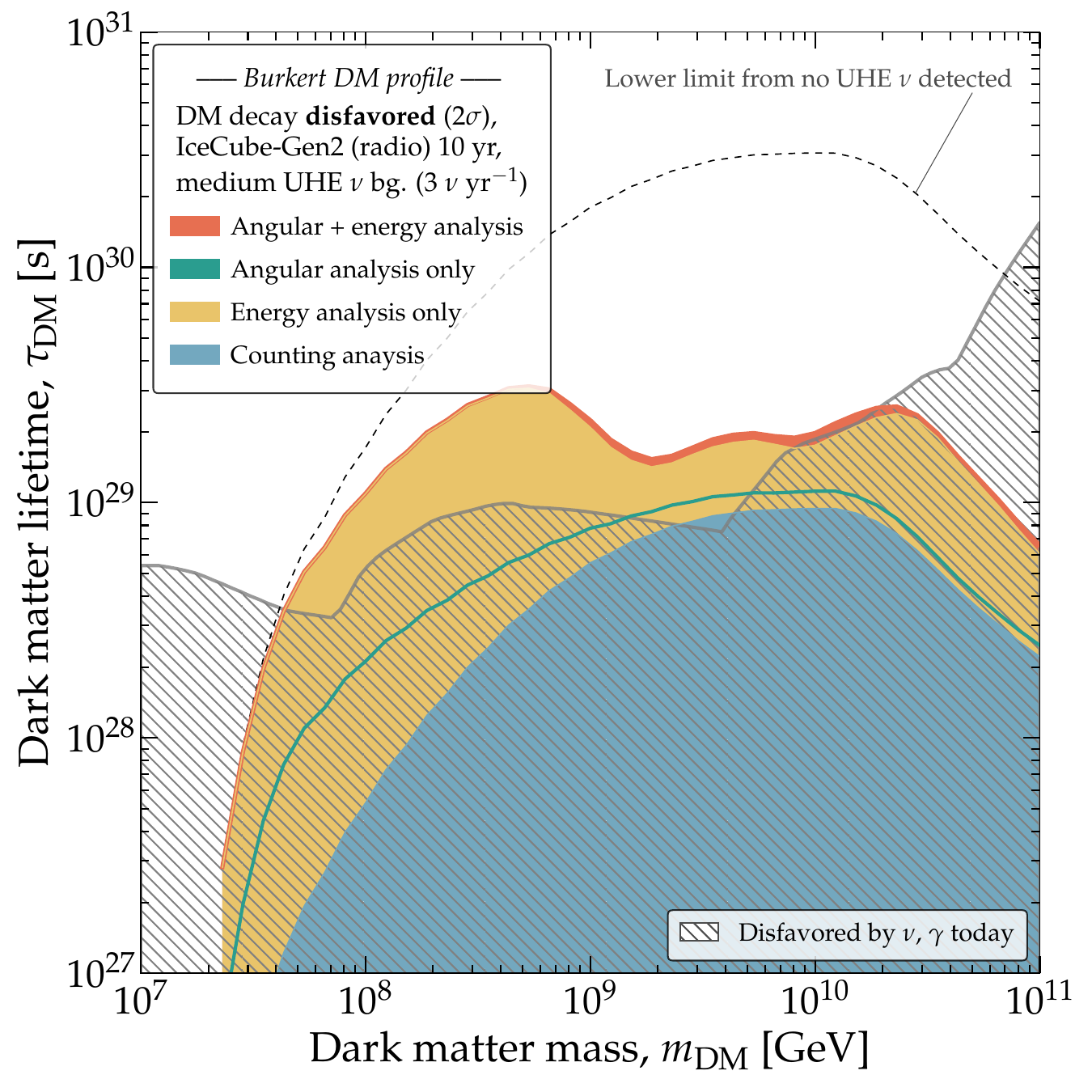}
 \includegraphics[width=0.49\textwidth]{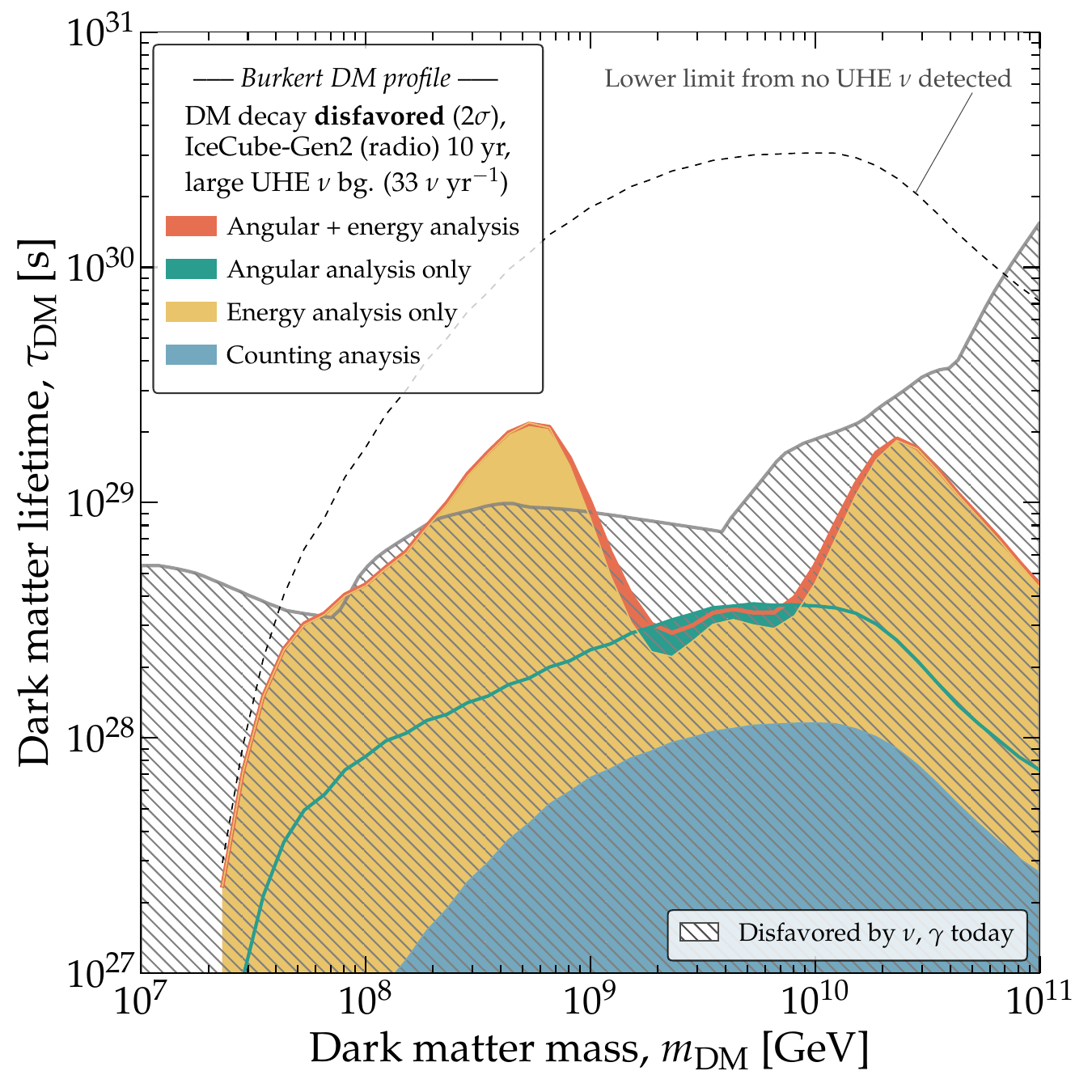}
 \caption{\textit{\textbf{Projected lower limits on the DM lifetime, using a Burkert Galactic DM profile.}  {\it Left:} For a medium background of non-DM UHE neutrinos, to be compared with \figu{bounds_medium_bg}.  {\it Right:} For a large background of non-DM UHE neutrinos, to be compared with \figu{bounds_large_bg}.}}\label{fig:bounds_burkert}
\end{figure*}

Without further ado, we include additional figures to complement those in the main text:
\begin{description}
 \item[Figure~\ref{fig:discovery_prospects_3sigma}]  Discovery prospects at $3\sigma$, for the NFW Galactic DM profile.
 \item[Figure~\ref{fig:bounds_large_bg}]  Lower limits on DM lifetime under a large background of non-DM UHE neutrinos, for the NFW profile.
 \item[Figure~\ref{fig:discovery_prospects_burkert}]  Discovery prospects at $\geq 2\sigma$ and $\geq 3\sigma$, for the Burkert Galactic DM profile.
 \item[Figure~\ref{fig:bounds_burkert}]  Lower limits on DM lifetime, for the Burkert Galactic DM profile.
\end{description}

\pagebreak
\newpage


\bibliography{refs}


\end{document}